\documentclass[aip,jcp,reprint,amsmath,amssymb,floatfix,citeautoscript]{revtex4-1}
\usepackage{cancel}
\usepackage{amsmath}
\usepackage{graphicx}
\usepackage{amssymb}
\usepackage{amsthm}
\usepackage{bm}
\usepackage{dcolumn}
\usepackage{braket}
\usepackage{longtable}
\usepackage{ragged2e}
\usepackage{txfonts}

\usepackage{color}
\usepackage[usenames,dvipsnames]{xcolor}
\definecolor{myblue}{rgb}{0,0,1}
\usepackage[breaklinks=true,colorlinks=true,linkcolor=myblue,urlcolor=myblue,citecolor=myblue]{hyperref}

\let\vr\undefined
\newcommand{\vr}{{\bm{r}}}

\newcommand{\vx}{{\bm{x}}}

\newcommand{\eps}{{\varepsilon}}

\newcommand{\nocc}{n_{\mathrm{occ}}}
\newcommand{\nvir}{n_{\mathrm{vir}}}

\begin{document}

\title{On The Relation Between Equation-of-Motion Coupled-Cluster Theory and the \textit{GW} Approximation}

\author{Malte F. Lange}
\affiliation{Department of Chemistry and James Franck Institute,
University of Chicago, Chicago, Illinois 60637, USA}

\author{Timothy C. Berkelbach}
\email{berkelbach@uchicago.edu}
\affiliation{Department of Chemistry and James Franck Institute,
University of Chicago, Chicago, Illinois 60637, USA}

\begin{abstract}

We discuss the analytic and diagrammatic structure of ionization potential (IP)
and electron affinity (EA) equation-of-motion coupled-cluster (EOM-CC) theory,
in order to put it on equal footing with the prevalent $GW$ approximation.  The
comparison is most straightforward for the time-ordered one-particle Green's
function, and we show that the Green's function calculated by EOM-CC with
single and double excitations (EOM-CCSD) includes fewer ring diagrams at higher
order than does the $GW$ approximation, due to the former's unbalanced treatment
of time-ordering.  However, the EOM-CCSD Green's function contains a large
number of vertex corrections, including ladder diagrams, mixed ring-ladder
diagrams, and exchange diagrams.  By including triple excitations, the EOM-CCSDT
Green's function includes all diagrams contained in the $GW$ approximation,
along with many high-order vertex corrections. In the same language, we discuss
a number of common approximations to the EOM-CCSD equations, many of which can
be classified as elimination of diagrams.  Finally, we present numerical results
by calculating the principal charged excitations energies of the molecules
contained in the so-called $GW$100 test set~[\textit{J.~Chem.~Theory
Comput.}~\textbf{2015}, \textit{11}, 5665--5687].  We argue that (in
molecules) exchange is as important as screening, advocating for a Hartree-Fock
reference and second-order exchange in the self-energy.

\vspace{1em}
\begin{centering}
\includegraphics[scale=1.0]{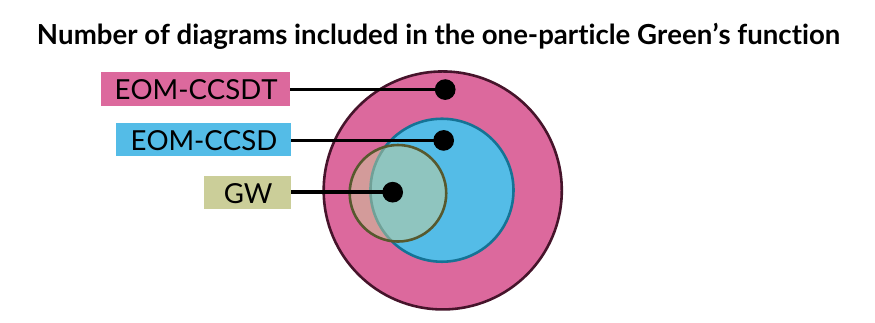}
\par
Table of contents figure.
\end{centering}

\end{abstract}

\maketitle

\section{Introduction}
\label{sec:intro}

The accurate calculation of excited-state properties constitutes one of the
major challenges in modern computational materials science.  For charged
excitations, namely the ionization potentials and electron affinities as
measured by photoelectron spectroscopy, the $GW$ approximation
has proven to be a powerful and successful tool in the condensed phase. 
Formally, the $GW$ approximation (reviewed below) arises as the lowest-order
self-energy diagram when the one-particle Green's function $G$ is expanded in
terms of the screened Coulomb interaction $W$~\cite{Hedin1965,Hybertsen1985},
with screening treated in the random-phase approximation.  Neglected diagrams
can be assigned to vertex corrections (appearing both in the self-energy and the
polarization propagator), which are a natural target for post-$GW$ theories and
an ongoing area of
activity~\cite{Schindlmayr1998,Shishkin2007,Romaniello2009,Gruneis2014,Schmidt2017}.

In contrast to time-dependent Green's function-based diagrammatic theories,
wavefunction-based theories and concomitant time-independent perturbation theory
offer an alternative route towards systematically improvable excited-state
calculations~\cite{ShavittBartlett}.  The great variety of wavefunction
ansatzes, combined with the long history of development and benchmarking in the
molecular quantum chemistry community, makes such approaches particularly
promising.  Unfortunately, the formal comparison between wavefunction-based and
Green's function-based techniques is complicated by a difference in both the
approach and the language.  Here, we present such a comparison, by analyzing the
one-particle Green's function calculated using equation-of-motion
coupled-cluster theory to that calculated using the $GW$ approximation.  A
relation between the two can be anticipated based on the known exact relation
between total \textit{ground-state} energies calculated using the ring
coupled-cluster doubles approach and using the random-phase
approximation~\cite{Freeman1977,RingSchuck,Scuseria2008}, the latter of which is at the
heart of screening in the $GW$ approximation.  However, for charged excitation
energies, the equivalence is not so straightforward.

Within the molecular physics and quantum chemistry communities, a number of
perturbative schemes have been proposed to directly construct the
self-energy~\cite{Cederbaum1975,vonNiessen1984}, including the outer-valence Green's function
approach~\cite{Cederbaum1975}, the two-particle-hole Tamm-Dancoff 
approximation~\cite{Schirmer1978} and its extended variant~\cite{Walter1981},
and the algebraic diagrammatic construction~\cite{Schirmer1983}.
More recently the second-order Green's function~\cite{SzaboOstlund,Holleboom1990} 
has been studied, especially
in its self-consistent~\cite{Phillips2014} and finite-temperature~\cite{Kananenka2016} variations.
In wavefunction-based techniques, ionization potentials and electron affinities
can either be calculated as a difference in ground-state energies (between the
neutral and ionic systems) or via the equation-of-motion framework, which
directly results in an eigensystem whose eigenvalues are the ionization potentials
or electron affinities. Equation-of-motion coupled-cluster (EOM-CC)
theory is one such framework, which typically achieves accurate excitation
energies when performed with single and double excitations
(EOM-CCSD)~\cite{Stanton1994,Krylov2008}.  At the intersection of these methods,
Nooijen and Snijders derived a one-particle Green's function in the CC
framework~\cite{Nooijen1992,Nooijen1993}, the poles and residues of which are
precisely those of the conventional EOM-CC formalism (in the bivariational
framework).  The CC Green's function has seen a renewed interest in recent
years~\cite{McClain2016,BhaskaranNair2016,Peng2016,Peng2018,Nishi2018}. One of
the main goals of the present work is to relate the latter theory to the $GW$
approximation, which is carried out in Sec.~\ref{sec:comp}.

A number of \textit{numerical} comparisons between Green's function-based and
wavefunction-based techniques for charged excitation energies have been carried
out in recent years.  In particular, comparisons between the $GW$ approximation
and wavefunction-based techniques have been performed for one-dimensional
lattice models~\cite{Ou2016}, for a test set of 24 organic acceptor
molecules~\cite{Knight2016}, for oligoacenes~\cite{Rangel2016}, and for a test
set of 100 molecules~\cite{Krause2015,Caruso2016}; the latter test set is known
as the $GW$100, introduced in Ref.~\onlinecite{vanSetten2015}, and forms the
basis of our numerical study in Sec.~\ref{sec:results}.

In light of recent efforts to bring the systematic improvability of
wavefunction-based theories into the solid
state~\cite{Marsman2009,Gruneis2010,Gruneis2011,DelBen2012,Gruneis2013,
Gruneis2015,Liao2016,McClain2016,Hummel2017,McClain2017},
we believe it timely to establish the relationship, both formally and
numerically, between popular wavefunction approaches and Green's function
approaches -- the latter of which has dominated solid-state electronic
structure.
Future work in both Green's function-based and wavefunction-based approaches can
benefit from the analysis and results of the present work.

The layout of this article is as follows.  In Sec.~\ref{sec:theory}, we provide
the requisite theoretical background associated with general features of the
one-particle Green's function, the $GW$ approximation to the self-energy,
and equation-of-motion coupled-cluster theory.  In Sec.~\ref{sec:comp},
we perform a detailed diagrammatic comparison of the two methods,
comparing separately their Green's functions and self-energies.  In Sec.~\ref{sec:results},
we use equation-of-motion coupled-cluster theory to calculate ionizations
potentials and electron affinities of the $GW$100 test set, and evaluate a number
of accurate but efficient approximations, which are straightforwardly analyzed 
with the previously introduced diagrammatic description.  In Sec.~\ref{sec:conc},
we conclude with an outlook for future developments.

\section{Theory}
\label{sec:theory}

\subsection{The one-particle Green's function}

The one-particle time-ordered Green's function is defined 
by~\cite{FetterWalecka,Cederbaum1975}
\begin{equation}
iG_{pq}(\omega) = \int d(t_1-t_2) e^{i\omega (t_1-t_2)} 
    \langle \Psi_0| T [\hat{a}_p(t_1) \hat{a}_q^\dagger(t_2)] |\Psi_0\rangle
\end{equation}
where $T$ is the time-ordering operator, $\Psi_0$ is the exact interacting
ground state, and $p,q$ index a complete set of single-particle spin-orbitals.
The irreducible self-energy matrix $\mathbf{\Sigma}(\omega)$ satisfies the
relation $\mathbf{G}(\omega) =
\left[\omega-\mathbf{f}-\mathbf{\Sigma(\omega)}\right]^{-1}$, where $\mathbf{f}$
is the matrix associated with some one-body (mean-field) operator such that
$\mathbf{\Sigma}$ contains the remaining effects of the electronic interactions.
Although in practice, $\mathbf{f}$ is commonly the Kohn-Sham matrix of density
functional theory, here we consider it to be the Fock matrix and will let
$p,q,r,s$ index the canonical Hartree-Fock (HF) orbitals, such that $\mathbf{f}$
is diagonal:~$f_{pq} = \varepsilon_p \delta_{pq}$.  Following convention,
indices $i,j,k,l$ are used for the $\nocc$ occupied orbitals in the HF
determinant and $a,b,c,d$ for the $\nvir$ virtual (unoccupied) orbitals; in
total there are $M = \nocc + \nvir$ orbitals.

\begin{figure}[b]
\centering
\includegraphics[scale=1.0]{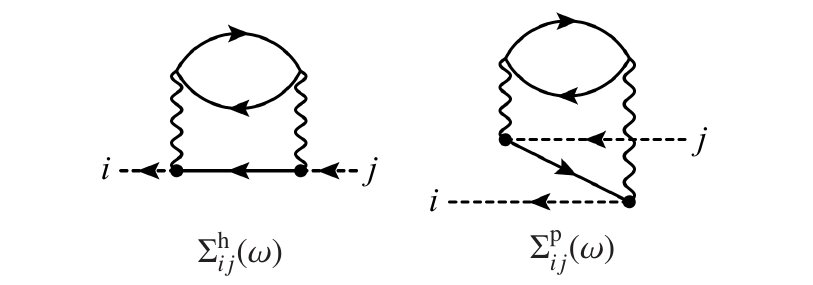}
\caption{
Lowest-order time-ordered (Goldstone) ring diagrams appearing in the hole~(h)
and particle~(p) contributions to the self-energy, for occupied orbitals $i,j$.
The dashed lines only serve to indicate the connectivity in a Goldstone
diagram for the Green's function.
}
\label{fig:sigma_time}
\end{figure}

\subsection{The \textit{GW} approximation}

The charged excitation energies (ionization potentials and electron affinities)
occur at the poles of the Green's function, i.e.~they are the self-consistent
eigenvalues of a frequency-dependent one-particle matrix $\mathbf{H}(\omega)$:
\begin{equation}
\sum_q H_{pq}(\omega=E_n) R^n_{q} = E_n R^n_p,
\end{equation} 
with
\begin{equation}
\label{eq:gw_heff}
H_{pq}(\omega) = \omega \delta_{pq} - [\mathbf{G}(\omega)]^{-1}_{pq} 
= \varepsilon_{p}\delta_{pq} + \Sigma_{pq}(\omega).
\end{equation}
In the $GW$ approximation~\cite{Hedin1965}, the self-energy is given by
\begin{equation}
\begin{split}
\Sigma(\vr_1,\vr_2;\omega) &= \frac{i}{2\pi}\int d\omega^\prime e^{i\eta \omega^\prime}
    G(\vr_1,\vr_2;\omega+\omega^\prime) \\
    &\hspace{10em} \times W_c(\vr_2,\vr_1;\omega^\prime),
\end{split}
\end{equation} 
where $W_c = W-v$ is the correlation part of the screened interaction;
recall that the bare exchange term has been included (self-consistently) in the
Fock operator $\mathbf{f}$.  The dielectric function
that screens the Coulomb interaction is evaluated with the random-phase
approximation (RPA), corresponding to a resummation of all ring diagrams
contributing to the polarization propagator; furthermore, all
vertex corrections are neglected.
Henceforth, we limit the discussion to the non-self-consistent $G_0W_0$
approximation, where $G_0$ and $W_0$ are evaluated in a ``one-shot'' manner
using the orbitals and orbital energies of the mean-field problem (in this case,
HF).  In a finite single-particle basis set, the frequency integration can be
done analytically to show that the self-energy has separate hole (h) and
particle (p) contributions~\cite{Pavlyukh2007,Bruneval2012,vanSetten2013},
\begin{align}
\label{eq:sigma_gw}
&\Sigma_{pq}(\omega)
    = \Sigma_{pq}^{\mathrm{h}}(\omega)
        + \Sigma_{pq}^{\mathrm{p}}(\omega) \\
&\hspace{0em} = \sum_\nu\left\{
        \sum_k \left[ \frac{[M_{kp}^\nu]^* M_{kq}^\nu}
            {\omega - (\varepsilon_k - \Omega_\nu) - i\eta} \right]
        + \sum_c \left[ \frac{[M_{cp}^\nu]^* M_{cq}^\nu}
            {\omega - (\varepsilon_c + \Omega_\nu) + i\eta} \right]
    \right\} \nonumber,
\end{align}
which are associated with the two possible time orderings (Goldstone diagrams)
of the corresponding Feynman diagram for the self-energy, i.e.~$\Sigma_{pq}(t_2-t_1)$
with $t_2 > t_1$ or with $t_2 < t_1$.  Expressed in terms of time-ordered
Goldstone diagrams, the lowest-order ring diagrams appearing
in the hole and particle contributions to the $GW$ self-energy are shown in
Fig.~\ref{fig:sigma_time}, for the single-particle indices $i,j$ in the occupied
orbital subspace.
The poles of the $GW$ self-energy occur at $\omega =
\varepsilon_k-\Omega_\nu$ and $\omega = \varepsilon_c + \Omega_\nu$, i.e.~at
sums and differences of the orbital energies $\varepsilon_p$ and neutral
excitation energies $\Omega_\nu$.
For future reference, we note that -- in gapped molecules and materials --
the particle contribution to the self-energy
of hole states is only weakly dependent on frequency, because
the quasiparticle energy $\omega \approx \varepsilon_k$
is far from the poles at $\varepsilon_c + \Omega_\nu$;
the separation is roughly twice the gap.  

The transition amplitudes associated with the poles of the self-energy are given by 
\begin{equation}
M^\nu_{pq} = \int d\vx_1 d\vx_2\ \rho_{\nu}(\vx_1) r_{12}^{-1}
    \phi_{p}^*(\vx_2) \phi_q(\vx_2) 
\end{equation}
where 
\begin{equation}
\rho_\nu(\vx) = \langle \Psi_0 | \hat{n}(\vx) |\Psi_\nu\rangle
    = \sum_{pq} \phi_p^*(\vx) \phi_q(\vx) \langle \Psi_0 | \hat{a}_p^\dagger \hat{a}_q | \Psi_\nu\rangle
\end{equation}
is the transition density of the neutral excited state $\Psi_\nu$.

The level of theory used to construct the polarizability 
determines the energies $\Omega_\nu$ and wavefunctions $\Psi_\nu$
entering in the above equations.  Using any time-dependent mean-field 
response, 
$|\Psi_\nu\rangle = \sum_{ai} \left[X_{ai}^\nu a_a^\dagger a_i 
    - Y_{ai}^\nu a_i^\dagger a_a \right] |\Psi_0\rangle$,
leads to an eigensystem commonly 
associated with the RPA, 
\begin{equation}
\label{eq:rpa}
\left(
\begin{array}{cc}
\mathbf{A}   & \mathbf{B} \\
-\mathbf{B}^* & -\mathbf{A}^*
\end{array}
\right)
\left(
\begin{array}{c}
\mathbf{X}^\nu \\
\mathbf{Y}^\nu
\end{array}
\right)
= \Omega_\nu
\left(
\begin{array}{c}
\mathbf{X}^\nu \\
\mathbf{Y}^\nu
\end{array}
\right),
\end{equation}
and transition amplitudes
\begin{equation}
\label{eq:gw_amps}
M^\nu_{pq} 
    = \sum_{ai} \left[ X^\nu_{ai} \langle ip | aq \rangle
    + Y^\nu_{ai} \langle ap | iq \rangle \right],
\end{equation}
where the two-electron integrals are given by
$\langle pq|rs\rangle \equiv \int d\bm{x}_1 d\bm{x}_2 
    \phi_p^*(\bm{x}_1) \phi_q^*(\bm{x}_2) r_{12}^{-1} \phi_r(\bm{x}_1) \phi_s(\bm{x}_2)$
and $\bm{x}$ is a combined spin and spatial variable.

Specifically using time-dependent HF theory, the $\mathbf{A}$ and 
$\mathbf{B}$ matrices (each of dimension $\nocc \nvir \times \nocc \nvir$) have elements
\begin{subequations}
\label{eq:RPAMatrices}
\begin{align}
A_{ai,bj} &= (\varepsilon_a-\varepsilon_i)\delta_{ab}\delta_{ij} + \langle aj||ib \rangle, \\
B_{ai,bj} &= \langle ij || ab \rangle. 
\end{align}
\end{subequations}
where the anti-symmetrized two-electron integrals are 
$\langle pq || rs \rangle \equiv \langle pq | rs\rangle - \langle pq |sr \rangle$.
Using the more conventional time-dependent Hartree dielectric function yields
the same structure, but neglects the exchange integrals in the $\mathbf{A}$ and
$\mathbf{B}$ matrices; this corresponds to the common version of the RPA and the one
used in the $GW$ approximation. 
The RPA eigenvalues come in positive- and negative-energy pairs, comprising
only $\nocc\nvir$ distinct eigenvalues; thus there are $M\nocc\nvir$ poles in the $GW$
self-energy.

In the form given here, the solution of the RPA eigenvalue problem in
Eq.~(\ref{eq:rpa}) highlights the canonical $N^6$ scaling of the $GW$
approximation~\cite{Bruneval2012,Bruneval2016}, which is identical to that of
EOM-CCSD.  This $GW$ scaling comes from the need to calculate \textit{all} RPA
eigenvalues in order to reliably calculate just one quasiparticle energy in
Eq.~(\ref{eq:sigma_gw}). Alternative formulations can reduce this scaling.

\subsection{Equation-of-motion coupled-cluster theory}
Equation-of-motion coupled-cluster theories start from the ground-state
CC wavefunction, $|\Psi\rangle = e^{\hat{T}}|\Phi\rangle$, where the cluster operator
$\hat{T}$ creates neutral excitations with respect to the reference determinant
$|\Phi\rangle$,
\begin{equation}
\label{eq:cc}
\hat{T} = \hat{T}_1 + \hat{T}_2 +\dots 
    = \sum_{ai} t_i^a \hat{a}^\dagger_a \hat{a}_i 
    + \frac{1}{4}\sum_{abij} t_{ij}^{ab} 
        \hat{a}_a^\dagger \hat{a}_b^\dagger \hat{a}_j \hat{a}_i
    + \dots
\end{equation}
The ground-state energy and cluster amplitudes are determined by the conditions 
\begin{subequations}
\begin{align}
E_{\mathrm{CC}} &= \langle \Phi | \bar{H} | \Phi \rangle \label{eq:gs} \\
0 &= \langle \Phi_i^a | \bar{H} | \Phi \rangle \label{eq:singles} \\
0 &= \langle \Phi_{ij}^{ab} | \bar{H} | \Phi \rangle \label {eq:doubles}
\end{align}
\end{subequations}
and so on, where 
$|\Phi_i^a\rangle = \hat{a}_a^\dagger \hat{a}_i |\Phi\rangle$, 
etc.~and $\bar{H} \equiv e^{-\hat{T}} \hat{H} e^{\hat{T}}$ is a similarity-transformed
Hamiltonian.  As seen above, the reference determinant is the
right-hand eigenvector of $\bar{H}$.  Because $\bar{H}$ is non-Hermitian, it has
distinct left-hand and right-hand eigenvectors for each eigenvalue; for the
ground state, the left-hand eigenvector of $\hat{H}$ is given by 
\begin{align}
\langle \tilde{\Psi}_0 | &= \langle \Phi | (1+\hat{\Lambda}) e^{-\hat{T}}, \\
\label{eq:lambda}
\hat{\Lambda} = \hat{\Lambda}_1 + \hat{\Lambda}_2 + \dots
    &= \sum_{ai} \lambda_{a}^{i} \hat{a}_i^\dagger \hat{a}_a 
    + \frac{1}{4} \sum_{abij} \lambda_{ab}^{ij} 
        \hat{a}_i^\dagger \hat{a}_j^\dagger \hat{a}_b \hat{a}_a + \dots
\end{align}

Charged excitation energies in EOM-CC are calculated as eigenvalues of a
$\bar{H}$ in a finite
basis of $(N\pm 1)$-electron Slater determinants; $(N-1)$-electron excitation
energies are calculated via the ionization potential (IP) framework and
$(N+1)$-electron excitation energies via the electron affinity (EA)
framework~\cite{Nooijen1992,Nooijen1993,Stanton1994,Krylov2008}.  For example,
the IP-EOM-CC energies are determined by
\begin{equation}
(E_{\mathrm{CC}}-\bar{H}) \hat{R}^{N-1}(n) |\Phi\rangle 
    = \Omega_n^{N-1} \hat{R}^{N-1}(n) |\Phi\rangle \label{eq:eom1}
\end{equation}
\begin{equation}
\begin{split}
\hat{R}^{N-1}(n) &= \hat{R}^{N-1}_1(n) + \hat{R}^{N-1}_2(n) + \dots \\
    & = \sum_{i} r_i(n) \hat{a}_i 
    + \frac{1}{2}\sum_{aij} r_{ij}^{a}(n) \hat{a}_a^\dagger \hat{a}_j \hat{a}_i
    + \dots \label{eq:eom2}
\end{split}
\end{equation}
where $\Omega_n^{N-1} = E_0^{N}-E_n^{N-1}$ is the negative of a many-body
ionization potential and corresponds to an exact pole of the one-particle
Green's function.  
Again, $\bar{H}$ has distinct left-hand eigenvectors,
\begin{equation}
\langle \Phi|\hat{L}^{N-1}(n)(E_{\mathrm{CC}}-\bar{H}) 
    = \langle \Phi|\hat{L}^{N-1}(n) \Omega_n^{N-1} \label{eq:eom1_left}
\end{equation}
\begin{equation}
\begin{split}
\hat{L}^{N-1}(n) &= \hat{L}^{N-1}_1(n) + \hat{L}^{N-1}_2(n) + \dots \\
&= \sum_{a} l^a(n) \hat{a}^\dagger_a 
    + \frac{1}{2}\sum_{abi} l_{i}^{ab}(n) \hat{a}_a^\dagger \hat{a}^\dagger_b \hat{a}_i
    + \dots \label{eq:eom2_left}
\end{split}
\end{equation}
The left-hand and right-hand eigenstates of the untransformed $\hat{H}$ are then given by 
\begin{align}
|\Psi_n^{N-1}\rangle &= e^{\hat{T}} \hat{R}^{N-1}(n) |\Phi\rangle \\
\langle \tilde{\Psi}_n^{N-1}| &= \langle \Phi|\hat{L}^{N-1}(n)e^{-\hat{T}}
\end{align}
and form a biorthogonal set.  With appropriate normalization, the eigenstates
yield a resolution-of-the-identity in the $(N\pm 1)$-electron space 
\begin{equation}
\begin{split}
1 &= \sum_n |\Psi_n^{N\pm 1}\rangle \langle \tilde{\Psi}_n^{N\pm 1}| \\
  &= \sum_n e^{\hat{T}}\hat{R}^{N\pm 1}(n)|\Phi\rangle 
        \langle \Phi|\hat{L}^{N\pm 1}(n)e^{-\hat{T}}.
\end{split}
\end{equation}
As first done by Nooijen and Snijders~\cite{Nooijen1992,Nooijen1993}, 
this enables an algebraic Lehmann representation
of the Green's function, which (as usual) separates into IP and EA
contributions due to the time-ordering operator,
$G_{pq}(\omega) = G_{pq}^{\mathrm{IP}}(\omega) + G_{pq}^{\mathrm{EA}}(\omega)$.
For example, the IP part is given by
\begin{subequations}
\begin{align}
G_{pq}^{\mathrm{IP}}(\omega) 
    &= \sum_n \frac{\tilde{\psi}_q(n) \psi_p(n)}{\omega - \Omega_n^{N-1} + i\eta}, \\
\tilde{\psi}_q(n) &= \langle \Phi | (1+\hat{\Lambda}) e^{-\hat{T}} 
        \hat{a}_q^\dagger e^{\hat{T}} \hat{R}^{N-1}(n) |\Phi\rangle, \\
\psi_p(n) &= \langle \Phi | \hat{L}^{N-1}(n) e^{-\hat{T}} \hat{a}_p e^{\hat{T}} |\Phi\rangle.
\end{align}
\end{subequations}

Using conventional many-body techniques
for the $\hat{T}$, $\hat{R}$, and $\hat{\Lambda}$ operators enables separate diagrammatic
expansions of the IP and EA contributions to the Green's
function~\cite{Nooijen1992,Nooijen1993}, which is properly size extensive as a sum of connected
diagrams.  In particular, using IP-EOM-CCSD, the IP Green's function is
given as the sum over all time-ordered (Goldstone) diagrams
for which cutting the diagram after each endpoint or vertex always leaves a sum of
disconnected diagrams at previous times, each of which has no more than two
electron and two hole open propagator lines.
Based on the outcome of this procedure, components of each diagram can be classified
as belonging to the cluster operators $\hat{T}_n$, the EOM operators
$\hat{R}_n$, or the de-excitation operators $\hat{\Lambda}_n$
(in Refs.~\onlinecite{Nooijen1992,Nooijen1993}, these operators are designated more
precisely as $\hat{T}_n$,
$\hat{S}^{(p)}_n(\omega)$, and $\hat{R}^{(pq)}_n(\omega)$ respectively).
Importantly in this construction, \textit{at each order in perturbation theory,
all time-orderings of a given Feynman diagram are not included}.

\section{Comparing the \textit{GW} approximation and EOM-CC theory}
\label{sec:comp}

\subsection{Comparing the Green's function}
\label{ssec:ccgf}

We first compare the time-ordered Goldstone diagrams appearing in the Green's
function of the $GW$ approximation and EOM-CC theory.  
By construction, the first-order terms in the Green's function are vanishing.
At second order, there are ten Feynman diagrams arising from six diagrams for
the proper self-energy, only two of which are not accounted for by a
self-consistent HF calculation.  The $GW$ Green's function includes only one of
these two diagrams, with a single ring, which translates to $4!=24$
Goldstone diagrams, \textit{all of which} are included in the EOM-CCSD Green's
function.  However, the EOM-CCSD Green's function also includes all Goldstone
diagrams associated with the second-order exchange diagram (another 24 Goldstone
diagrams).  Therefore, as is well known, the EOM-CCSD Green's function is
correct through second order, and thus exact for two-electron problems; the $GW$
Green's function is not. 

\begin{figure}[t]
\centering
\includegraphics[scale=1.0]{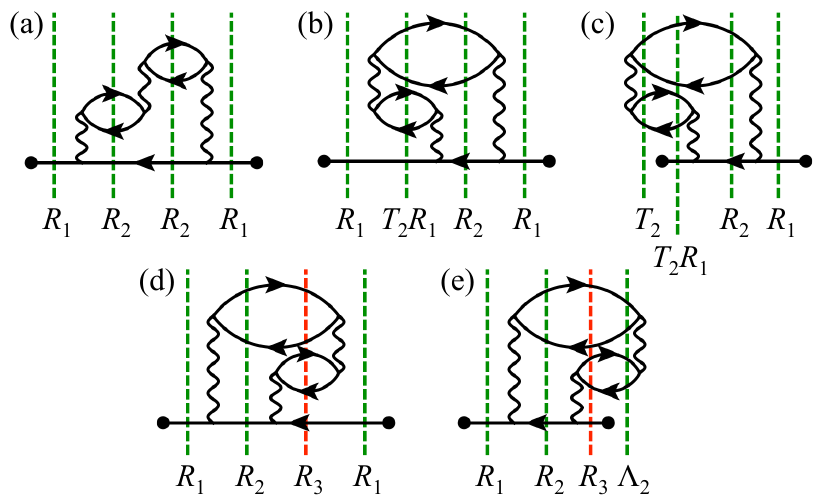}
\caption{The only five third-order Goldstone diagrams contributing to
$G^{\mathrm{IP}}_{ij}$ originating from $\Sigma_{ij}(t_1,t_2)$ with $t_1 < t_2$, 
in the $GW$ approximation.
Only (a), (b), and (c) are included in the EOM-CCSD Green's function; all five
are included in the EOM-CCSDT Green's function.  Time increases from left to
right.
}
\label{fig:gw_cc_third}
\end{figure}

At third order, the comparison is more complicated.  Again, the $GW$ Green's
function contains one irreducible Feynman diagram, which includes two rings
leading to $5! = 120$ Goldstone diagrams.  For simplicity of analysis,
we focus on IP diagrams ($t_1<t_2$)
in the occupied orbital subspace, $G^{\mathrm{IP}}_{ij}(t_1,t_2)$, generated
by the hole part of the self-energy, i.e.~$\Sigma_{ij}(t_1,t_2)$ with $t_1 < t_2$. 
The $GW$ approximation produces five such Goldstone diagrams, \textit{only three of
which} are included in the EOM-CCSD Green's function.  When cut after the second
interaction, diagrams (d) and (e) produce, at earlier times, a connected diagram
in the 3-hole+2-particle (3h2p) space, which is included in the EOM-CCSDT
Green's function, but not the EOM-CCSD one.  In contrast, diagrams (b) and (c)
also have a 3h2p configuration, but one that is generated by the disconnected
product of the $T_2$~[2-hole+2-particle (2h2p)] and $R_1$~[1-hole (1h)]
operators.  Therefore, \textit{some of the non-TDA $GW$ diagrams are included in the EOM-CCSD
Green's function, but not all of them.}

\begin{figure}[b]
\centering
\includegraphics[scale=1.0]{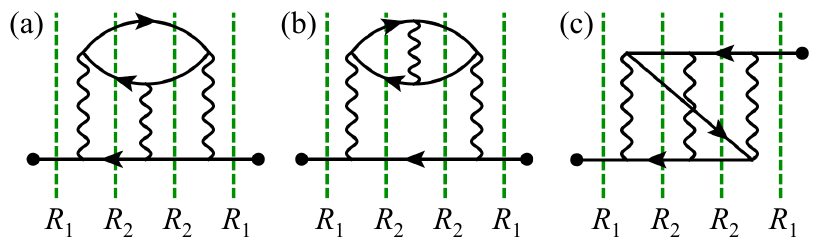}
\caption{Three example third-order Goldstone diagrams contributing to the IP
part of the EOM-CCSD Green's function with $t_1 < t_2$, which are not included
in the $GW$ Green's function.  All diagrams shown are generated by the EOM (2h1p)
formalism, independent of coupled-cluster theory.
Time increases from left to right.
}
\label{fig:cc_third_ladder}
\end{figure}

The above analysis is straightforward to generalize to higher order, and we find
that the irreducible part of the EOM-CCSD Green's function at $n$th order
contains only a vanishing fraction of the ring diagrams included in the $GW$
Green's function; at $n$th order, the fraction of diagrams included is $O(1/n)$.
(Of course, a large number of \textit{reducible} Green's function
diagrams are included at $n$th order, due to combinations of low-order
diagrams.) In spite of this apparent flaw of EOM-CCSD theory, we emphasize 
that the EOM-CCSD Green's function contains many other non-ring diagrams
that are not contained in the $GW$ approximation.  For example, three
third-order diagrams corresponding to various particle-particle and
particle-hole ladders are shown in Fig.~\ref{fig:cc_third_ladder}.
Diagrams (a) and (c) include vertex corrections to the self-energy
and diagram (b) includes a vertex correction to the polarization propagator.

The behavior we have described should be compared to the enumeration of
Goldstone diagrams for the correlation energy (the vacuum amplitude): in this
context, ground-state CCSD includes ring diagrams with \textit{all possible time
orderings}, completely encompassing those diagrams contained in the
RPA~\cite{Freeman1977,Scuseria2008}.  Similarly, one of us (T.C.B.) has recently
shown that the frequency-dependent polarizability calculated with
neutral-excitation EOM-CCSD encompasses all diagrams contained in the
(dynamical) RPA~\cite{Emrich1981,Berkelbach2018}.
The difference observed here for the one-particle Green's function can be traced
to the need for the EOM-CC operators to simultaneously describe screening and
free-particle propagation, as exemplified in diagrams (d) and (e) in
Fig.~\ref{fig:gw_cc_third}.  For exactly this reason, the CCSD correlation
energy \textit{is not recovered} from the EOM-CCSD Green's function, as
discussed by Nooijen and Snijders~\cite{Nooijen1993},
\begin{equation}
\label{eq:ecorr}
E_{\mathrm{CCSD}} \neq \frac{1}{4\pi i} \int d\omega 
    \mathrm{Tr} \left\{(\mathbf{h}+\omega\mathbf{1})\mathbf{G}^{\mathrm{IP}}_{\mathrm{CCSD}}(\omega)\right\}.
\end{equation}
Because of Eq.~(\ref{eq:ecorr}), there is no EOM-ring-CCD Green's function that
produces the RPA correlation energy.
The above can roughly be viewed as a reminder that although the CCSD energy
is exact to third order in perturbation theory, the EOM-CCSD energies are only
exact to second order.  
However, the EOM-CCSD Green's function does yield the CCSD reduced density matrix
and is thus properly number-conserving, 
\begin{align}
\bm{\rho}_{\mathrm{CCSD}} &= \frac{1}{2\pi i}\int d\omega [\mathbf{G}^{\mathrm{IP}}_{\mathrm{CCSD}}(\omega)] \\
N &= \mathrm{Tr}\left\{ \bm{\rho}_{\mathrm{CCSD}} \right\}.
\end{align}
Therefore, despite the error in the individual poles of the EOM-CCSD Green's function,
some ``sum rules'' are satisfied.

In this section, we have compared the Green's functions generated by EOM-CC and
the $GW$ approximation.  A more direct connection with the $GW$ approximation
and related time-dependent diagrammatic methods can be made by directly
targeting an EOM-CC self-energy or polarization propagator; work along these
lines is currently in progress in our group.  However, an \textit{approximate}
algebraic self-energy can be worked out directly from the EOM-CC eigenvalue
problem, which we turn to next.

\subsection{Comparing the self-energy}

For the remainder of the article, we will only consider IP-EOM-CCSD; the results
for EA-EOM-CCSD are completely analogous.
We introduce the normal-ordered Hamiltonian, with respect to the HF reference,
$\hat{H}_\mathrm{N} \equiv \hat{H} - E_{\mathrm{HF}}$ and its similarity-transformed variant
$\bar{H}_\mathrm{N} \equiv \bar{H} - E_{\mathrm{CC}}$.
The linear eigenvalue problem in Eqs.~(\ref{eq:eom1}) and (\ref{eq:eom2}) clearly leads
to a (schematic) matrix representation
\begin{equation}
\mathbf{\Delta\bar{H}}_\mathrm{CC} =
-\left(
\begin{array}{cc}
\langle \Phi_i | \bar{H}_\mathrm{N} |\Phi_k \rangle & \langle \Phi_i |\bar{H}_\mathrm{N}| \Phi_{kl}^{b} \rangle \\
\langle \Phi_{ij}^{a} | \bar{H}_\mathrm{N} | \Phi_{k} \rangle & \langle \Phi_{ij}^{a} |\bar{H}_\mathrm{N}| \Phi_{kl}^{b} \rangle
\end{array}
\right); \label{eq:eom_matrix}
\end{equation}
In this section, we will show that the $GW$ excitation energies are closely related to
the eigenvalues of the approximated matrix
\begin{equation}
\mathbf{\Delta\bar{H}}_{\mathrm{CC}GW} =
-\left(
\begin{array}{cc}
\langle \Phi_i | e^{-\hat{T}_2}\hat{f}_\mathrm{N} e^{\hat{T}_2} |\Phi_k \rangle 
    & \langle \Phi_i |\bar{H}_\mathrm{N}| \Phi_{kl}^{b} \rangle \\
\langle \Phi_{ij}^{a} | \bar{H}_\mathrm{N} | \Phi_{k} \rangle 
    & \langle \Phi_{ij}^{a} | H_{\mathrm{N}} | \Phi_{kl}^{b} \rangle
\end{array}
\right), \label{eq:ccgw}
\end{equation}
where $\hat{f}_\mathrm{N}$ is the normal-ordered Fock operator, $\hat{T}_1 = 0$
everywhere, the $\hat{T}_2$ amplitudes satisfy an approximate version of
Eq.~(\ref{eq:doubles}) known as ``ring-CCD''~\cite{Freeman1977,Scuseria2008},
and the untransformed Hamiltonian ($\hat{T}_2 = 0$) is used in the doubles-doubles
block.  

If desired, antisymmetrization can further be removed from most two-electron
integrals leading to the use of ``direct
ring-CCD''~\cite{Freeman1977,Scuseria2008} in the one-hole space combined with a
more conventional TDH treatment of screening in the two-hole+one-particle.
However, we keep antisymmetrization throughout, which makes the theory
manifestly self-interaction free, while retaining only the essential ingredients
of the $GW$ approximation.

Using a L\"owdin partitioning~\cite{Lowdin1963}, the eigenvalues of the Hamiltonian in
Eq.~(\ref{eq:ccgw}) can be found self-consistently for the
frequency-dependent matrix
\begin{equation}
\label{eq:eom_lowdin}
\begin{split}
A_{ij}(\omega) &= -\langle \Phi_i | e^{-\hat{T}_2} \hat{f}_\mathrm{N} e^{\hat{T}_2} | \Phi_j \rangle \\
    &\hspace{1em} + \frac{1}{4} \sum_{abklmn} \langle \Phi_{i} | \bar{H} | \Phi_{kl}^{a} \rangle 
    [G_{\mathrm{2h1p}}]_{klmn}^{ab}(\omega) \langle \Phi_{mn}^{b} | \bar{H} | \Phi_j\rangle \\
&= \varepsilon_{i}\delta_{ij} + \tilde{\Sigma}_{ij}^{\mathrm{p}} + \tilde{\Sigma}_{ij}^{\mathrm{h}}(\omega)
\end{split}
\end{equation}
where $G_{\mathrm{2h1p}}(\omega)$ is a specific time-sequence of the three-particle Green's function,
\begin{equation}
\begin{split}
[G_\mathrm{2h1p}]_{klmn}^{ab}(\omega) 
    &= -i\int dt e^{i\omega t} \langle \Phi_0 | [\hat{a}_a^\dagger \hat{a}_k \hat{a}_l](0) 
        [\hat{a}_b^\dagger \hat{a}_m \hat{a}_n](t) | \Phi_0\rangle \\
    &= \langle \Phi_0 | \hat{a}_a^\dagger \hat{a}_k \hat{a}_l 
        \frac{1}{\omega - (-\mathcal{P} \hat{H}_\mathrm{N}\mathcal{P} )} 
        \hat{a}_b^\dagger \hat{a}_m \hat{a}_n |\Phi_0\rangle; 
\end{split}
\end{equation}
this Green's function describes propagation in the 2h1p subspace
generated by the projection operator $\mathcal{P}$.
The self-consistent eigenvalue problem defined in Eq.~(\ref{eq:eom_lowdin}) is analogous
to that of the $GW$ approximation defined in Eqs.~(\ref{eq:gw_heff}) and (\ref{eq:sigma_gw}).
There are two significant differences that originate from the treatment of
time-ordering in EOM-CC theory.  First, the particle contribution to the
self-energy of the IP part of the Green's function
$\tilde{\Sigma}_{ij}^{\mathrm{p}}$ is frequency-independent; analogously, in
EA-EOM-CC, the hole contribution to the self-energy of the EA part of the
Green's function is frequency-independent.  However, as discussed above, the
frequency dependence of these terms -- when calculating the respective
excitation energy -- is typically very weak.  Second, the effective self-energy
in IP-EOM-CC (resp.~EA-EOM-CC) only has matrix elements in the occupied
(virtual) orbital space; this is in contrast to the self-energy in any proper
diagrammatic theory, which has matrix elements in the entire orbital space.  We
emphasize that neither of these differences reflects an approximation, but only
a difference in formalism; diagrammatically-defined self-energy theories and
EOM-CC can both be made exact in their appropriate limits, while retaining their
respective (different) mathematical structures.

\begin{figure*}[t]
\centering
\includegraphics[scale=1.0]{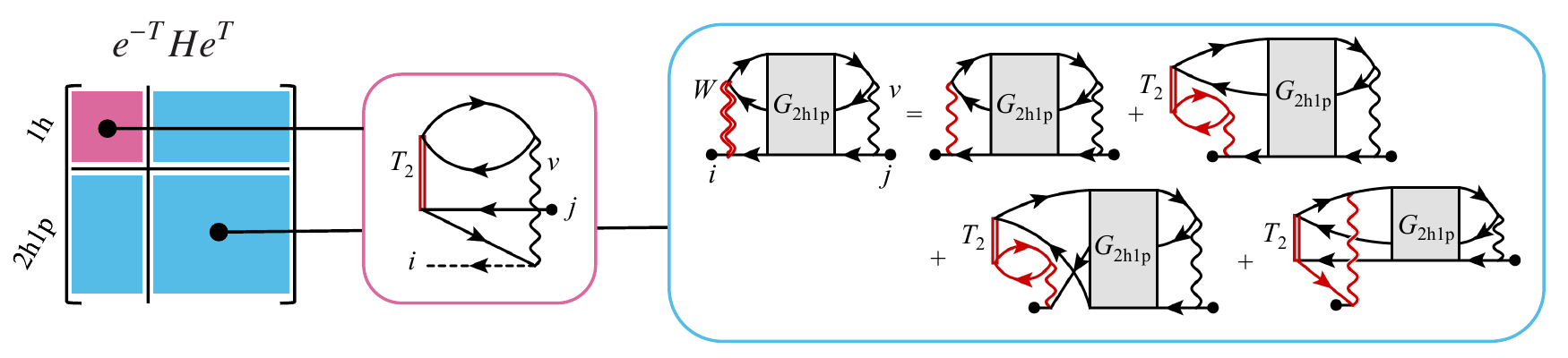}
\caption{
Goldstone self-energy diagrams arising from the different sectors of the
IP-EOM-CCSD similarity-transformed Hamiltonian.
The $\hat{T}_2$-transformed Fock operator in the 1h space generates a frequency-independent
screened-exchange self-energy $\tilde{\Sigma}_{ij}^{\mathrm{p}}$.  
The $\hat{T}_2$-transformed repulsion integral
$W_{iakl}$, which couples the 1h and 2h1p spaces, generates non-Tamm-Dancoff screening
on top of the Tamm-Dancoff ring and ladder diagrams generated in the 2h1p space; this leads
to a frequency-dependent screened-exchange self-energy $\tilde{\Sigma}_{ij}^{\mathrm{h}}(\omega)$.
}
\label{fig:cc_heff}
\end{figure*}

To summarize the structure of this effective ``self-energy'' from an
EOM-CC-based theory: for the IPs, the forward time-ordered self-energy (the hole
contribution) arises from coupling between 1h and 2h1p configurations, whereas
the reverse time-ordered self-energy (the particle contribution) arises from the
similarity transformation of the Fock operator in the 1h subspace; this behavior
is shown schematically in Fig.~\ref{fig:cc_heff}.  In the $GW$
language, both of these effects can be viewed as giving rise to screening of the
quasiparticle excitations.  From this point of view, we observe that the
IP-EOM-CISD methodology~\cite{Golubeva2009}, obtained by setting $\hat{T}_1 = \hat{T}_2 = 0$
in IP-EOM-CCSD, only includes one of the two time orderings, each diagram of
which is fully forward-time-ordered in the TDA sense.
Finally, we note that the \textit{true} CCSD self-energy can be straightforwardly
obtained numerically, by calculating both the IP and EA Green's functions and using
Dyson's equation,
\begin{equation}
\Sigma_{pq}(\omega) = (\omega - \varepsilon_{p})\delta_{pq} 
    - \left[ \mathbf{G}^{\mathrm{IP}}(\omega) 
    + \mathbf{G}^{\mathrm{EA}}(\omega) \right]^{-1}_{pq}.
\end{equation}
Naturally, the frequency-dependent matrix $\omega - \mathbf{f} - \mathbf{\Sigma}(\omega)$
will have eigenvalues given exactly by the IP/EA-EOM-CCSD excitation energies,
as well as the proper analytical (frequency-dependent) structure.

We now proceed to make the comparison between the approximate IP-EOM-CCSD 
of Eq.~(\ref{eq:ccgw}) and the $GW$ approximation more explicit.
First, we consider the frequency-dependent hole contribution
$\Sigma_{ij}^{\mathrm{h}}(\omega)$. 
The 2h1p Green's function can be expressed in two ways: first, as a
perturbative series that can be translated into diagrams,
\begin{equation}
\begin{split}
[G_\mathrm{2h1p}]_{klmn}^{ab}(\omega) 
    &= \sum_{p=0}^{\infty}\langle \Phi_{kl}^{a} |\hat{R}_0(\omega) 
        \left[\mathcal{P}\hat{V}_\mathrm{N} \hat{R}_0(\omega)\right]^p |\Phi_{mn}^{b}\rangle \\
\end{split}
\end{equation}
where $\hat{R}_0(\omega) = \left[\omega - (-\hat{f}_\mathrm{N})\right]^{-1}$ is the resolvent
of the Fock operator and 
$\hat{V}_\mathrm{N} = \hat{H}_\mathrm{N} - \hat{f}_\mathrm{N}$ is the normal-ordered fluctuation
operator.
Alternatively, $G_\mathrm{2h1p}$ can be expressed in terms of the solutions of an eigenproblem,
\begin{equation}
[G_\mathrm{2h1p}]_{klmn}^{ab}(\omega) 
    = \sum_{\nu} \frac{r_{kl}^{a}(\nu) [r_{mn}^{b}(\nu)]^*}
    {\omega-E^{\mathrm{2h1p}}_\nu} 
\end{equation}
where $E_{\nu}^{\mathrm{2h1p}}$ is an eigenvalue of the 2h1p block
of the Hamiltonian $\langle \Phi_{kl}^{a} | (-\bar{H}_\mathrm{N}) | \Phi_{mn}^{b} \rangle$
with eigenvector 
$|\nu\rangle = \frac{1}{2}\sum_{akl} r_{kl}^{a}(\nu)|\Phi_{kl}^{a}\rangle$.
Notably, the set of diagrams contained in $G_\mathrm{2h1p}$ is identically those
included in the two-particle-hole TDA theory of the
self-energy~\cite{Schirmer1978}, mentioned in the introduction.  The CC
self-energy goes beyond the TDA diagrams via the outer vertices, i.e.~the
matrix elements of the similarity-transformed Hamiltonian, which can be
evaluated to give
\begin{subequations}
\begin{align}
\label{eq:wovoo}
\begin{split}
\langle \Phi_i | \bar{H} | \Phi_{kl}^{a} \rangle &= W_{iakl} 
    = \langle ia || kl \rangle 
    + \sum_{me} \langle im||ke\rangle t_{lm}^{ae}, \\
    &\hspace{1em} - \sum_{me} \langle im||le\rangle t_{km}^{ae} 
    + \frac{1}{2} \sum_{ef} \langle ia||ef\rangle t_{kl}^{ef}
\end{split} \\
\langle \Phi_{mn}^{b} | \bar{H} | \Phi_j \rangle &= \langle mn || jb \rangle,
\end{align}
\end{subequations}
leading to the self-energy
\begin{equation}
\label{eq:cc_hole}
\tilde{\Sigma}^{\mathrm{h}}_{ij}(\omega)
    = \frac{1}{4} \sum_{abklmn} W_{iakl}
    [G_\mathrm{2h1p}]_{klmn}^{ab}(\omega) \langle mn || jb \rangle.
\end{equation}
Viewing $W$ as a screened Coulomb interaction leads to the set of diagrams shown
in Fig.~\ref{fig:cc_heff}.  The construction of the intermediate $W$ has a
non-iterative $N^6$ cost, which is usually swamped by the iterative $N^5$ cost of
subsequent matrix-vector multiplies during Davidson diagonalization.

The use of exact CCSD amplitudes in Eq.~(\ref{eq:wovoo}) includes many beyond-$GW$
insertions in the polarization propagator.  However, as discussed, the closest
comparison can be made when the $\hat{T}_2$ amplitudes solve the approximate ring-CCD
equations,
\begin{equation}
\begin{split}
&t_{ij}^{ab}(\varepsilon_i+\varepsilon_j-\varepsilon_a-\varepsilon_b)
  = \langle ab || ij \rangle \\
  &\hspace{1em} + \sum_{ck} t_{ik}^{ac} \langle kb || cj \rangle
  + \sum_{ck} \langle ak || ic \rangle t_{kj}^{cb}
  + \sum_{cdkl} t_{ik}^{ac}\langle kl || cd \rangle t_{lj}^{db}. \label{eq:ringccd}
\end{split}
\end{equation}
Iteration of these equations adds higher-order non-TDA ring diagrams in the
self-energy, very much like in the $GW$ approximation.  However, consistent with
the analysis presented in Sec.~\ref{ssec:ccgf}, \textit{the non-TDA diagrams are
generated in an asymmetric and incomplete manner}.

Beyond this issue of non-TDA diagrams on the later-time side of the self-energy,
the approximation described so far has three additional qualitative differences
from the $GW$ approximation.  First, the presence of antisymmetrized vertices
generates many exchange diagrams not included in the conventional $GW$
approximation.  In particular, the ``exterior'' antisymmetrization is
responsible for some of the self-energy diagrams that are in the second-order
screened exchange (SOSEX) approach~\cite{Ren2015} and ``interior''
antisymmetrization yields particle-hole ladders that improve the quality of the
polarization propagator.  Second, the 2h1p Green's function includes the
interaction between two holes in the intermediate 2h1p state, leading to
hole-hole ladder insertions, which are vertex corrections beyond the structure
of the $GW$ self-energy.  Third, the final term in Eq.~(\ref{eq:wovoo}) can be
shown to produce mixed ring-ladder diagrams that are not included at the $GW$ or
SOSEX levels of theory.

Finally, we consider the frequency-independent particle contribution to the
self-energy. In IP-EOM-CCD, this term is given by
\begin{equation}
\tilde{\Sigma}_{ij}^\mathrm{p} = \frac{1}{2} \sum_{kab} \langle ik||ab\rangle t_{kj}^{ba},
\label{eq:cc_part}
\end{equation}
which can be represented by the single diagram shown in Fig.~\ref{fig:cc_heff}.
This diagram must be evaluated as a scalar without frequency dependence
according to the usual diagrammatic rules of time-independent perturbation
theory~\cite{ShavittBartlett}. With this interpretation, the iteration of the
ring-CCSD equations again generates all 
TDA-screening diagrams plus an asymmetric subset of non-TDA screening diagrams.
Antisymmetrization is responsible for subsets of both $GW$ and SOSEX diagrams.

\section{Application of EOM-CCSD to the $GW$100 test set}
\label{sec:results}

Having established the formal relation between EOM-CCSD and the $GW$
approximation, we now present a numerical comparison.  In particular, we will
study the so-called $GW$100 test set~\cite{vanSetten2015}, comprising 100 small- to medium-sized
molecules with up to 66 active electrons in 400 spatial orbitals.  The $GW$100
test set was introduced by van Setten and co-authors~\cite{vanSetten2015} in
order to provide a simple and controlled class of problems with which to compare
theoretical and computational approximations of $GW$-based implementations.
This important research agenda aims to enforce reproducibility within the
community and highlight the successes and limitations of the aforementioned
time-dependent diagrammatic techniques, thereby identifying avenues for future
research.  The $GW$100 has been studied by a number of different 
groups~\cite{vanSetten2015,Krause2015,Caruso2016,Maggio2017,Govoni2018}.

In addition to providing results and analysis for conventional IP- and 
EA-EOM-CCSD excitation energies, we will also consider a number of
approximations.  These approximations make the computational cost more
competitive with that of the $GW$ approximation and -- in light of the
previous sections -- many of them can be understood as selective inclusion
of certain diagrams.
These approximations are described in the next section.  
Some of these approximations have been investigated and compared for
charged excitations~\cite{Dutta2013,Dutta2014,Dutta2018}
and for neutral electronic excitations~\cite{Goings2014,Rishi2017}.

\subsection{Approximations to EOM-CCSD}

As mentioned previously, in their canonical forms, the $GW$ approximation and
the IP/EA-EOM-CCSD formalism both scale as $N^6$.  For the latter class of methods,
this scaling originates from the solution of the ground-state CCSD equations,
while the subsequent ionized EOM eigenvalue problem exhibits only $N^5$ scaling
(with relatively cheap, non-iterative $N^6$ steps associated with construction of the
intermediates).  For this reason, a natural target for approximations
leading to reduced cost is the ground-state calculation.  Despite the
distinction we draw between ground-state and excited-state approximations, we
note that the results of the previous section have shown that the determination
of the $T$-amplitudes via ground-state CCSD directly affects the diagrams
contributing to the one-particle Green's function.

\textit{MBPT2 ground state}.
The most severe approximation to the ground state is that of second-order
many-body perturbation theory (MBPT2). For a canonical Hartree-Fock reference,
which we use throughout this work, this is equivalent to second-order M\o
ller-Plesset perturbation theory (MP2).  In this approach, $\hat{T}_1=0$
and the $\hat{T}_2$ amplitudes are approximated by 
\begin{equation}
t_{ij}^{ab} \approx \frac{\langle ab || ij \rangle}{\eps_i + \eps_j - \eps_a - \eps_b}.
\end{equation}
Due to the transformation from atomic orbitals to molecular orbitals, an MBPT2
calculation scales as $N^5$ and so the use of MBPT2 amplitudes in an IP/EA-EOM
calculation leads to overall $N^5$ methods for ionization
potentials and electron affinities.  Following
Refs.~\onlinecite{Nooijen1995,Gwaltney1996,Goings2014}, we call this method
EOM-MBPT2; the same method has also been referred to as
EOM-CCSD(2)~\cite{Stanton1995}.

\textit{CC2 ground state}.
A popular approximation to reduce the cost of CCSD is the CC2
model~\cite{Christiansen1995}.  In this technique, the $\hat{T}_1$ amplitude equations
are unchanged from those of CCSD, while the $\hat{T}_2$ amplitude
equations~(\ref{eq:doubles}) are changed such that $\hat{T}_2$ only connects to
the Fock operator
\begin{equation}
0 = \langle \Phi_{ij}^{ab} | e^{-\hat{T}_1}\hat{H}_\mathrm{N}e^{\hat{T}_1} 
    + e^{-\hat{T}_2}\hat{f}_\mathrm{N}e^{\hat{T}_2} | \Phi \rangle.
\end{equation}
This leads to approximate $\hat{T}_2$ amplitudes that are very similar to those of MBPT2,
\begin{equation}
t_{ij}^{ab} \approx \frac{\overline{\langle ab||ij\rangle}}{\eps_i+\eps_j-\eps_a-\eps_b},
\end{equation}
where
$\overline{\langle ab||ij\rangle} 
    \equiv \langle \Phi_{ij}^{ab} | e^{-\hat{T}_1} \hat{H}_\mathrm{N}e^{\hat{T}_1} |\Phi\rangle$
are $\hat{T}_1$-transformed two-electron integrals.
Like MBPT2, the CC2 approximation removes the $\hat{T}_2$ contractions responsible for $N^6$
scaling and is thus an iterative $N^5$ technique.
While CC2 treats dynamical correlation at essentially the same level as MBPT2, the full
treatment of single excitations generated by $\hat{T}_1$ allows orbital relaxation, which
should be beneficial in cases where the HF determinant is suboptimal.

\textit{Linearized CCSD}.
The final ground-state approximation that we consider is linearized CCSD
(linCCSD)~\cite{Byrd2015}, which is the least severe approximation to CCSD.  In
this approach, all quadratic products of the CCSD amplitudes are neglected in
the amplitude equations.  In diagrammatic language, this approximation neglects
many -- \textit{but not all} -- of the non-TDA diagrams in the Green's function;
for example, the third-order non-TDA diagrams shown in
Fig.~\ref{fig:gw_cc_third}(b) and (c) are included even when the amplitude
equations are linearized.  The non-TDA time-ordering is a result of the
combination of $\hat{R}_1$ and $\hat{T}_2$, rather than of nonlinear terms in
the $\hat{T}_2$ equations.  Although linearized CCSD still scales as $N^6$, the
method is more amenable to parallelization~\cite{Byrd2015}, which may be
desirable for large systems or
solids~\cite{Booth2013,Liao2016,McClain2017,Hummel2017}.

\textit{Excited state approximation}.
After the ground-state calculation, the most expensive contribution to an
EOM-CCSD calculation comes from the large doubles-doubles block of the
similarity-transformed Hamiltonian.  
A natural approximation then is to replace the doubles-doubles block by simple
orbital-energy differences,
\begin{equation}
\langle \Phi_{ij}^{a} | \bar{H}_{\mathrm{N}} | \Phi_{kl}^{b} \rangle
    \approx (\varepsilon_a - \varepsilon_i - \varepsilon_j)\delta_{ab}\delta_{ik}\delta_{jl},
\end{equation}
leading to a diagonal structure and a straightforward L\"owdin partitioning.
Naturally, this approach is only reasonable for principle charged excitations
with a large weight in the singles (one-hole or one-particle) sector.

This partitioned variant of EOM-CCSD theory still exhibits $N^5$ scaling after the
$\hat{T}$-amplitudes are determined, but requires the construction and storage of far
fewer integral intermediates.
Formally, this approximate partitioning technique can be combined with any
treatment of the ground-state CC equations, though it only makes practical
sense for approximate ground-state calculations whose cost does not overwhelm
that of the EOM calculation.  We will combine the approximate partitioning
technique with MBPT2 and CC2 ground states, denoting the results as
P-EOM-MBPT2 and P-EOM-CC2, respectively.

As first discussed in 
Ref.~\onlinecite{Nooijen1995}, the P-EOM-MBPT2
method is formally very close to the non-self-consistent
second-order Green's function technique (GF2), where the self-energy
is composed of second-order ring and exchange
diagrams~\cite{SzaboOstlund,NegeleOrland}.  When applied exactly as described,
P-EOM-MBPT2 actually includes a few third-order self-energy diagrams --
as can be seen in Fig.~\ref{fig:cc_heff}.  These can be removed by
also neglecting the $\hat{T}$-amplitudes in the screened Coulomb interaction
$W_{iakl}$ that couples the 1h and 2h1p space, given in Eq.~(\ref{eq:wovoo}).
However, this additional ground-state correlation is found to be responsible for
a remarkable improvement in the accuracy of P-EOM-MBPT2 when compared to GF2.

With respect to the hierarchy of linear-response CC2 methods described in
Ref.~\onlinecite{Walz2016}, the EOM-CC2 method described here is equivalent to
IP-CCSD[f]$_{\mathrm{CC2}}$ and the P-EOM-CC2 method is between
IP-CCSD[0]$_{\mathrm{CC2}}$ and IP-CCSD[1]$_{\mathrm{CC2}}$.

\subsection{Numerical details}
\label{sec:results}

We have applied the above methods to calculate the first few principle
ionization potentials and electron affinities for the molecules in the $GW$100
test set.  Followings Refs.~\onlinecite{Krause2015,Caruso2016}, we work in the
localized-orbital def2-TZVPP basis set~\cite{Weigend2005}, using corresponding
pseudopotentials for elements in the fifth and sixth row of the periodic table;
core orbitals were frozen in all calculations.  While this choice of basis is a
good trade-off between cost and accuracy, our results are not converged with
respect to the basis set and should not be compared directly to experiment or to
calculations in other basis sets, such as plane-wave based $GW$
calculations~\cite{Maggio2017,Govoni2018}.  Instead, these calculations can be
directly compared to preexisting ionization potentials in the same
basis~\cite{Krause2015,Caruso2016}.  More importantly, our calculations are
internally consistent; the main purpose of this section is to benchmark the
accuracy of cost-saving approximations to EOM-CCSD and demonstrate the utility
of EOM-CCSD techniques for excited-state properties of benchmark data sets.
Extrapolation to the complete basis set limit and comparison with other non-CC
techniques is reserved for future work.

An advantage of the IP/EA-EOM-CCSD approaches is an avoidance of open-shell
calculations for charged molecules.  As such, all of our calculations were performed
using a spin-free implementation based on a closed-shell restricted HF reference,
and free of spin contamination.
All calculations were performed using the PySCF software package~\cite{Sun2018}.

Recent work~\cite{Maggio2017} has identified two molecules from the original $GW$100
test set with incorrect geometries: vinyl bromide and phenol. For consistency with
previously published results, we have performed calculations on the original
geometries. 

\subsection{Comparison to $\Delta$CCSD(T) ionization potentials}

\begin{figure}[t]
\centering
\includegraphics[scale=1.0]{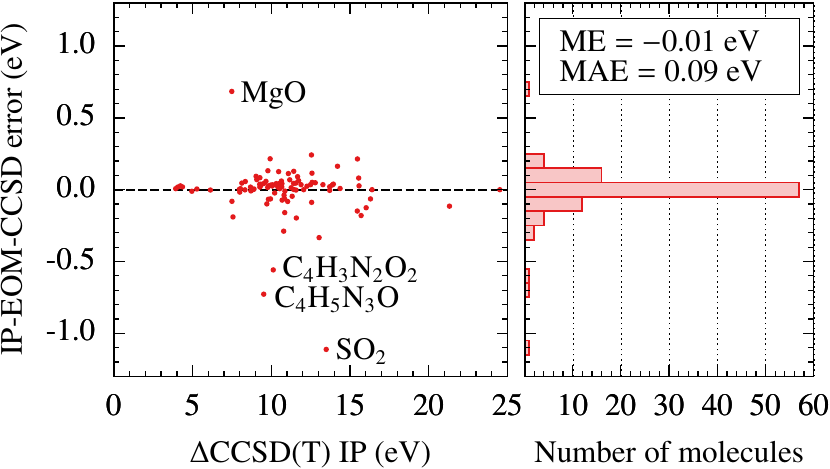}
\caption{Comparison of ionization potentials predicted by IP-EOM-CCSD
compared to those predicted by $\Delta$CCSD(T) from Ref.~\onlinecite{Krause2015}.
Errors with respect to the latter are presented as a scatter plot (left) and
histogram (right). Four molecules with errors exceeding 0.5~eV are explicitly
labeled.}
\label{fig:ccsdt}
\end{figure}

We first aim to establish the accuracy of EOM-CCSD for the $GW$100 test set.  As
a ground-state theory, CCSD with perturbative triples [CCSD(T)] represents the
`gold standard' for weakly-correlated medium-sized molecules~\cite{Bartlett2007} and scales as
$N^7$, which is more expensive than any method considered here.  At this high level
of theory, the first IP of each molecule has been calculated by Krause et
al.~\cite{Krause2015}, as a difference in ground-state energies between neutral
and charged molecules -- the so-called $\Delta$CCSD(T) scheme.  Higher-energy
IPs and EAs, in particular those with the same symmetry as the first, cannot be
calculated using this approach 

In Fig.~\ref{fig:ccsdt}, we show the comparison between IPs predicted by
$\Delta$CCSD(T) and IP-EOM-CCSD.  The IP-EOM-CCSD values exhibit a signed mean
error (ME) of $-0.01$~eV and a mean absolute error (MAE) of 0.09~eV.  The small
mean error indicates that the errors are not systematic.  Only four
molecules, identified in Fig.~\ref{fig:ccsdt} have errors larger than 0.5~eV,
suggesting that IP-EOM-CCSD represents a good approximation to $\Delta$CCSD(T),
at least for the molecules included in the $GW$100.  The most fair comparison,
based on diagrams generated, is to the $G_0W_0$ approximation based on a HF
reference ($G_0W_0$@HF).  As reported by Caruso et al.~\cite{Caruso2016}, such an approach
overestimates IPs, leading to a ME of 0.26~eV and a MAE of 0.35~eV.
Interestingly, this is significantly better than the more popular (at least in
the solid state) $G_0W_0$ approximation based on a PBE starting point
($G_0W_0$@PBE), which
severely and systematically underestimates IPs, leading to a ME of $-0.69$~eV and a MAE of
0.69~eV~\cite{Caruso2016}.

\subsection{Electron affinities and higher-energy excitations}

By construction, EOM-CCSD can straightforwardly predict higher-energy ionization
potentials, corresponding to more deeply-bound electrons, as well as the first
and higher electron affinities.  For completeness, in Tab.~\ref{tab:eomccsd} we
report the first three occupied and unoccupied quasiparticle energies (i.e.~the
negative of the first three IPs and EAs with large quasiparticle weights) for 
each molecule in the $GW$100 test
set, as calculated by IP- and EA-EOM-CCSD, and accounting for their
multiplicities.

\begin{longtable*}{@{\extracolsep{\fill}} l l r r r r r r }
\hline\hline
Formula & Name  & HOMO-2  & HOMO-1  & HOMO  & LUMO  & LUMO+1  & LUMO+2   \\
\hline
He  &  helium   &  & &  $-24.51$ $(\times 1)$&  $22.22$ $(\times 1)$ &  $39.82$ $(\times 3)$ &  $166.90$ $(\times 1)$  \\
Ne  &  neon   & &  $-48.33$ $(\times 1)$&  $-21.21$ $(\times 3)$ &  $20.84$ $(\times 1)$ &  $21.87$ $(\times 2)$ &  $74.27$ $(\times 3)$  \\
Ar  &  argon   & &  $-29.58$ $(\times 1)$&  $-15.63$ $(\times 2)$ &  $14.73$ $(\times 3)$ &  $17.19$ $(\times 4)$ &  $20.69$ $(\times 1)$  \\
Kr  &  krypton   & &  $-27.14$ $(\times 1)$&  $-13.98$ $(\times 3)$ &  $10.41$ $(\times 2)$ &  $12.24$ $(\times 4)$ &  $18.78$ $(\times 1)$  \\
Xe  &  xenon   & &  $-23.67$ $(\times 1)$&  $-12.23$ $(\times 3)$ &  $7.72$ $(\times 3)$ &  $8.91$ $(\times 3)$ &  $12.29$ $(\times 1)$  \\
H$_2$  &  hydrogen   &  & &  $-16.40$ $(\times 1)$&  $4.22$ $(\times 1)$ &  $8.05$ $(\times 1)$ &  $16.04$ $(\times 1)$  \\
Li$_2$  &  lithium dimer  &  $-63.39$ $(\times 1)$&  $-63.38$ $(\times 1)$ &  $-5.27$ $(\times 1)$ &  $-0.12$ $(\times 1)$ &  $0.99$ $(\times 2)$ &  $1.00$ $(\times 1)$  \\
Na$_2$  &  sodium dimer  &  $-37.24$ $(\times 2)$ &  $-37.21$ $(\times 1)$ &  $-4.94$ $(\times 1)$ &  $-0.26$ $(\times 1)$ &  $0.60$ $(\times 2)$ &  $0.78$ $(\times 1)$  \\
Na$_4$  &  sodium tetramer  &  $-36.60$ $(\times 1)$&  $-5.59$ $(\times 1)$ &  $-4.25$ $(\times 1)$ &  $-0.53$ $(\times 1)$ &  $-0.10$ $(\times 1)$ &  $0.10$ $(\times 1)$  \\
Na$_6$  &  sodium hexamer  &  $-36.52$ $(\times 1)$&  $-5.68$ $(\times 1)$ &  $-4.37$ $(\times 2)$ &  $-0.49$ $(\times 1)$ &  $-0.34$ $(\times 2)$ &  $-0.07$ $(\times 1)$  \\
K$_2$  &  potassium dimer  &  $-23.83$ $(\times 2)$ &  $-23.77$ $(\times 1)$ &  $-4.08$ $(\times 1)$ &  $-0.32$ $(\times 1)$ &  $0.77$ $(\times 1)$ &  $0.83$ $(\times 2)$  \\
Rb$_2$  &  rubidium dimer  &  $-20.09$ $(\times 2)$ &  $-20.02$ $(\times 1)$ &  $-3.93$ $(\times 1)$ &  $-0.37$ $(\times 1)$ &  $0.18$ $(\times 1)$ &  $0.35$ $(\times 1)$  \\
N$_2$  &  nitrogen  &  $-18.84$ $(\times 1)$ &  $-17.21$ $(\times 1)$ &  $-15.60$ $(\times 1)$ &  $3.05$ $(\times 2)$ &  $8.97$ $(\times 1)$ &  $9.47$ $(\times 1)$  \\
P$_2$  &  phosphorus dimer  &  $-14.79$ $(\times 1)$ &  $-10.75$ $(\times 1)$ &  $-10.59$ $(\times 1)$ &  $-0.10$ $(\times 2)$ &  $3.47$ $(\times 1)$ &  $6.30$ $(\times 1)$  \\
As$_2$  &  arsenic dimer  &  $-14.64$ $(\times 1)$ &  $-10.14$ $(\times 1)$ &  $-9.91$ $(\times 2)$ &  $-0.26$ $(\times 1)$ &  $3.31$ $(\times 1)$ &  $6.88$ $(\times 1)$  \\
F$_2$  &  fluorine  &  $-21.09$ $(\times 1)$ &  $-18.85$ $(\times 1)$ &  $-15.53$ $(\times 1)$ &  $0.39$ $(\times 1)$ &  $15.34$ $(\times 1)$ &  $15.34$ $(\times 1)$  \\
Cl$_2$  &  chlorine  &  $-15.98$ $(\times 1)$ &  $-14.41$ $(\times 1)$ &  $-11.46$ $(\times 2)$ &  $-0.19$ $(\times 1)$ &  $9.10$ $(\times 1)$ &  $10.03$ $(\times 1)$  \\
Br$_2$  &  bromine  &  $-14.44$ $(\times 1)$ &  $-12.89$ $(\times 1)$ &  $-10.54$ $(\times 1)$ &  $-0.94$ $(\times 1)$ &  $7.05$ $(\times 1)$ &  $7.09$ $(\times 1)$  \\
I$_2$  &  iodine  &  $-12.76$ $(\times 1)$ &  $-11.38$ $(\times 2)$ &  $-9.55$ $(\times 2)$ &  $-1.50$ $(\times 1)$ &  $5.10$ $(\times 1)$ &  $5.15$ $(\times 1)$  \\
CH$_4$  &  methane   & &  $-23.38$ $(\times 1)$&  $-14.38$ $(\times 3)$ &  $3.45$ $(\times 1)$ &  $5.79$ $(\times 3)$ &  $7.88$ $(\times 3)$  \\
C$_2$H$_6$  &  ethane  &  $-13.05$ $(\times 1)$ &  $-12.71$ $(\times 1)$ &  $-12.71$ $(\times 1)$ &  $3.11$ $(\times 1)$ &  $4.21$ $(\times 1)$ &  $5.24$ $(\times 2)$  \\
C$_3$H$_8$  &  propane  &  $-12.25$ $(\times 1)$ &  $-12.12$ $(\times 1)$ &  $-12.05$ $(\times 1)$ &  $2.95$ $(\times 1)$ &  $4.13$ $(\times 1)$ &  $4.32$ $(\times 1)$  \\
C$_4$H$_{10}$  &  butane  &  $-11.92$ $(\times 1)$ &  $-11.81$ $(\times 1)$ &  $-11.56$ $(\times 1)$ &  $2.88$ $(\times 1)$ &  $3.46$ $(\times 1)$ &  $4.12$ $(\times 1)$  \\
C$_2$H$_4$  &  ethlyene  &  $-14.91$ $(\times 1)$ &  $-13.11$ $(\times 1)$ &  $-10.69$ $(\times 1)$ &  $2.63$ $(\times 1)$ &  $3.94$ $(\times 1)$ &  $4.60$ $(\times 1)$  \\
C$_2$H$_2$  &  ethyne  &  $-19.13$ $(\times 1)$ &  $-17.23$ $(\times 1)$ &  $-11.55$ $(\times 2)$ &  $3.50$ $(\times 2)$ &  $3.58$ $(\times 1)$ &  $4.53$ $(\times 1)$  \\
C$_4$  &  tetracarbon  &  $-14.65$ $(\times 1)$ &  $-11.46$ $(\times 1)$ &  $-11.27$ $(\times 1)$ &  $-2.36$ $(\times 1)$ &  $-0.01$ $(\times 1)$ &  $1.71$ $(\times 1)$  \\
C$_3$H$_6$  &  cyclopropane  &  $-13.14$ $(\times 1)$ &  $-10.86$ $(\times 1)$ &  $-10.85$ $(\times 1)$ &  $3.46$ $(\times 1)$ &  $3.96$ $(\times 1)$ &  $4.10$ $(\times 1)$  \\
C$_6$H$_6$  &  benzene  &  $-12.14$ $(\times 2)$ &  $-9.32$ $(\times 1)$ &  $-9.32$ $(\times 1)$ &  $1.78$ $(\times 2)$ &  $3.11$ $(\times 1)$ &  $4.00$ $(\times 2)$  \\
C$_8$H$_8$  &  cyclooctatetraene  &  $-10.01$ $(\times 1)$ &  $-10.00$ $(\times 1)$ &  $-8.40$ $(\times 1)$ &  $0.79$ $(\times 1)$ &  $2.52$ $(\times 1)$ &  $2.54$ $(\times 1)$  \\
C$_5$H$_6$  &  cyclopentadiene  &  $-12.52$ $(\times 1)$ &  $-11.03$ $(\times 1)$ &  $-8.69$ $(\times 1)$ &  $1.77$ $(\times 1)$ &  $3.27$ $(\times 1)$ &  $4.06$ $(\times 1)$  \\
C$_2$H$_3$F  &  vinyl fluoride  &  $-14.79$ $(\times 1)$ &  $-13.86$ $(\times 1)$ &  $-10.60$ $(\times 1)$ &  $2.80$ $(\times 1)$ &  $3.93$ $(\times 1)$ &  $4.34$ $(\times 1)$  \\
C$_2$H$_3$Cl  &  vinyl chloride  &  $-13.21$ $(\times 1)$ &  $-11.65$ $(\times 1)$ &  $-10.13$ $(\times 1)$ &  $2.12$ $(\times 1)$ &  $3.49$ $(\times 1)$ &  $3.84$ $(\times 1)$  \\
C$_2$H$_3$Br  &  vinyl bromide  &  $-13.38$ $(\times 1)$ &  $-10.71$ $(\times 1)$ &  $-9.29$ $(\times 1)$ &  $2.02$ $(\times 1)$ &  $3.55$ $(\times 1)$ &  $4.27$ $(\times 1)$  \\
C$_2$H$_3$I  &  vinyl iodide  &  $-11.71$ $(\times 1)$ &  $-9.92$ $(\times 1)$ &  $-9.36$ $(\times 1)$ &  $1.40$ $(\times 1)$ &  $1.75$ $(\times 1)$ &  $3.59$ $(\times 1)$  \\
CF$_4$  &  tetrafluoromethane  &  $-18.35$ $(\times 2)$ &  $-17.37$ $(\times 2)$ &  $-16.24$ $(\times 3)$ &  $4.89$ $(\times 1)$ &  $6.86$ $(\times 2)$ &  $9.20$ $(\times 2)$  \\
CCl$_4$  &  tetrachloromethane  &  $-13.40$ $(\times 2)$ &  $-12.46$ $(\times 3)$ &  $-11.60$ $(\times 3)$ &  $0.86$ $(\times 1)$ &  $2.20$ $(\times 3)$ &  $5.20$ $(\times 1)$  \\
CBr$_4$  &  tetrabromomethane  &  $-12.13$ $(\times 2)$ &  $-11.25$ $(\times 3)$ &  $-10.48$ $(\times 2)$ &  $-0.49$ $(\times 1)$ &  $1.21$ $(\times 3)$ &  $4.60$ $(\times 1)$  \\
CI$_4$  &  tetraiodomethane   & &  $-9.99$ $(\times 2)$&  $-9.30$ $(\times 2)$ &  $-1.62$ $(\times 1)$ &  $0.34$ $(\times 2)$ &  $4.73$ $(\times 2)$ \\
SiH$_4$  &  silane   & &  $-18.46$ $(\times 1)$&  $-12.84$ $(\times 3)$ &  $3.10$ $(\times 3)$ &  $3.71$ $(\times 1)$ &  $6.75$ $(\times 2)$  \\
GeH$_4$  &  germane  &  $-38.07$ $(\times 1)$&  $-18.69$ $(\times 1)$ &  $-12.53$ $(\times 2)$ &  $3.16$ $(\times 1)$ &  $3.54$ $(\times 3)$ &  $7.04$ $(\times 2)$  \\
Si$_2$H$_6$  &  disilane  &  $-12.25$ $(\times 1)$ &  $-12.25$ $(\times 1)$ &  $-10.71$ $(\times 1)$ &  $2.27$ $(\times 1)$ &  $2.28$ $(\times 2)$ &  $2.75$ $(\times 1)$  \\
Si$_5$H$_{12}$  &  pentasilane  &  $-10.84$ $(\times 1)$ &  $-10.64$ $(\times 1)$ &  $-9.36$ $(\times 1)$ &  $0.79$ $(\times 1)$ &  $1.57$ $(\times 1)$ &  $1.61$ $(\times 1)$  \\
LiH  &  lithium hydride   & &  $-64.54$ $(\times 1)$&  $-7.96$ $(\times 1)$ &  $0.09$ $(\times 1)$ &  $2.01$ $(\times 2)$ &  $3.41$ $(\times 1)$  \\
KH  &  potassium hydride  &  $-24.59$ $(\times 2)$ &  $-24.38$ $(\times 1)$ &  $-6.13$ $(\times 1)$ &  $-0.04$ $(\times 1)$ &  $1.60$ $(\times 2)$ &  $1.90$ $(\times 1)$  \\
BH$_3$  &  borane   & &  $-18.35$ $(\times 1)$&  $-13.31$ $(\times 2)$ &  $0.33$ $(\times 1)$ &  $3.36$ $(\times 1)$ &  $4.32$ $(\times 2)$  \\
B$_2$H$_6$  &  diborane  &  $-14.00$ $(\times 1)$ &  $-13.48$ $(\times 1)$ &  $-12.29$ $(\times 1)$ &  $1.20$ $(\times 1)$ &  $2.51$ $(\times 1)$ &  $3.46$ $(\times 1)$  \\
NH$_3$  &  ammonia  &  $-27.78$ $(\times 1)$&  $-16.52$ $(\times 2)$ &  $-10.77$ $(\times 1)$ &  $2.84$ $(\times 1)$ &  $5.26$ $(\times 2)$ &  $11.18$ $(\times 2)$  \\
HN$_3$  &  hydrazoic acid  &  $-15.92$ $(\times 1)$ &  $-12.25$ $(\times 1)$ &  $-10.72$ $(\times 1)$ &  $2.02$ $(\times 1)$ &  $3.02$ $(\times 1)$ &  $3.17$ $(\times 1)$  \\
PH$_3$  &  phosphine  &  $-20.25$ $(\times 1)$&  $-13.75$ $(\times 2)$ &  $-10.57$ $(\times 1)$ &  $2.95$ $(\times 1)$ &  $3.12$ $(\times 2)$ &  $7.11$ $(\times 2)$  \\
AsH$_3$  &  arsine  &  $-19.82$ $(\times 1)$ &  $-13.18$ $(\times 2)$ &  $-10.42$ $(\times 1)$ &  $2.86$ $(\times 1)$ &  $3.01$ $(\times 2)$ &  $7.44$ $(\times 2)$  \\
SH$_2$  &  hydrogen sulfide  &  $-15.65$ $(\times 1)$ &  $-13.39$ $(\times 1)$ &  $-10.35$ $(\times 1)$ &  $2.79$ $(\times 1)$ &  $3.20$ $(\times 1)$ &  $7.25$ $(\times 1)$  \\
FH  &  hydrogen fluoride  &  $-39.30$ $(\times 1)$&  $-19.84$ $(\times 1)$ &  $-15.90$ $(\times 1)$ &  $3.07$ $(\times 1)$ &  $14.20$ $(\times 1)$ &  $17.24$ $(\times 1)$  \\
ClH  &  hydrogen chloride  &  $-25.44$ $(\times 1)$&  $-16.65$ $(\times 1)$ &  $-12.64$ $(\times 1)$ &  $2.70$ $(\times 1)$ &  $7.91$ $(\times 1)$ &  $12.18$ $(\times 1)$  \\
LiF  &  lithium fluoride  &  $-33.11$ $(\times 1)$ &  $-11.76$ $(\times 1)$ &  $-11.28$ $(\times 2)$ &  $-0.02$ $(\times 1)$ &  $2.74$ $(\times 2)$ &  $3.51$ $(\times 1)$  \\
F$_2$Mg  &  magnesium fluoride  &  $-14.15$ $(\times 1)$ &  $-13.76$ $(\times 2)$ &  $-13.71$ $(\times 1)$ &  $-0.04$ $(\times 1)$ &  $1.94$ $(\times 2)$ &  $4.13$ $(\times 1)$  \\
TiF$_4$  &  titanium fluoride  &  $-16.86$ $(\times 1)$ &  $-16.29$ $(\times 2)$ &  $-15.69$ $(\times 2)$ &  $-1.06$ $(\times 2)$ &  $0.07$ $(\times 3)$ &  $0.98$ $(\times 1)$  \\
AlF$_3$  &  aluminum fluoride  &  $-16.08$ $(\times 2)$ &  $-15.86$ $(\times 2)$ &  $-15.31$ $(\times 1)$ &  $0.67$ $(\times 1)$ &  $1.86$ $(\times 1)$ &  $3.90$ $(\times 2)$  \\
BF  &  boron monofluoride  &  $-21.27$ $(\times 1)$ &  $-18.16$ $(\times 2)$ &  $-11.20$ $(\times 1)$ &  $1.51$ $(\times 2)$ &  $3.29$ $(\times 1)$ &  $4.70$ $(\times 1)$  \\
SF$_4$  &  sulfur tetrafluoride  &  $-15.26$ $(\times 1)$ &  $-15.04$ $(\times 1)$ &  $-12.70$ $(\times 1)$ &  $0.94$ $(\times 1)$ &  $3.79$ $(\times 1)$ &  $4.86$ $(\times 1)$  \\
BrK  &  potassium bromide  &  $-19.37$ $(\times 1)$ &  $-8.41$ $(\times 1)$ &  $-8.17$ $(\times 2)$ &  $-0.45$ $(\times 1)$ &  $1.40$ $(\times 2)$ &  $1.73$ $(\times 1)$  \\
GaCl  &  gallium monochloride  &  $-14.09$ $(\times 1)$ &  $-11.46$ $(\times 2)$ &  $-9.79$ $(\times 1)$ &  $0.27$ $(\times 2)$ &  $2.49$ $(\times 1)$ &  $6.60$ $(\times 1)$  \\
NaCl  &  sodium chloride  &  $-20.83$ $(\times 1)$ &  $-9.55$ $(\times 1)$ &  $-9.12$ $(\times 2)$ &  $-0.59$ $(\times 1)$ &  $1.18$ $(\times 2)$ &  $2.08$ $(\times 1)$  \\
MgCl$_2$  &  magnesium chloride  &  $-12.52$ $(\times 1)$ &  $-11.88$ $(\times 2)$ &  $-11.76$ $(\times 2)$ &  $-0.19$ $(\times 1)$ &  $1.40$ $(\times 2)$ &  $3.91$ $(\times 1)$  \\
AlI$_3$  &  aluminum chloride  &  $-10.39$ $(\times 2)$ &  $-10.29$ $(\times 2)$ &  $-9.84$ $(\times 1)$ &  $-0.33$ $(\times 1)$ &  $-0.28$ $(\times 1)$ &  $2.25$ $(\times 2)$  \\
BN  &  boron nitride  &  $-27.96$ $(\times 1)$&  $-13.69$ $(\times 1)$ &  $-11.93$ $(\times 2)$ &  $-3.16$ $(\times 1)$ &  $2.65$ $(\times 2)$ &  $3.69$ $(\times 1)$  \\
NCH  &  hydrogen cyanide  &  $-20.60$ $(\times 1)$ &  $-13.91$ $(\times 1)$ &  $-13.90$ $(\times 1)$ &  $3.21$ $(\times 2)$ &  $3.45$ $(\times 1)$ &  $4.55$ $(\times 1)$  \\
PN  &  phosphorus mononitride  &  $-16.37$ $(\times 1)$ &  $-12.42$ $(\times 1)$ &  $-11.80$ $(\times 1)$ &  $0.46$ $(\times 1)$ &  $3.40$ $(\times 1)$ &  $8.42$ $(\times 1)$  \\
H$_2$NNH$_2$  &  hydrazine  &  $-15.39$ $(\times 1)$ &  $-11.28$ $(\times 1)$ &  $-9.62$ $(\times 1)$ &  $2.51$ $(\times 1)$ &  $3.69$ $(\times 1)$ &  $4.55$ $(\times 1)$  \\
H$_2$CO  &  formaldehyde  &  $-16.04$ $(\times 1)$ &  $-14.56$ $(\times 1)$ &  $-10.78$ $(\times 1)$ &  $1.67$ $(\times 1)$ &  $3.68$ $(\times 1)$ &  $5.24$ $(\times 1)$  \\
CH$_4$O  &  methanol  &  $-14.32$ $(\times 1)$ &  $-13.33$ $(\times 1)$ &  $-10.18$ $(\times 1)$ &  $1.91$ $(\times 1)$ &  $3.14$ $(\times 1)$ &  $4.00$ $(\times 1)$  \\
C$_2$H$_6$O  &  ethanol  &  $-13.43$ $(\times 1)$ &  $-12.27$ $(\times 1)$ &  $-10.61$ $(\times 1)$ &  $2.86$ $(\times 1)$ &  $3.73$ $(\times 1)$ &  $4.69$ $(\times 1)$  \\
C$_2$H$_4$O  &  acetaldehyde  &  $-14.32$ $(\times 1)$ &  $-13.33$ $(\times 1)$ &  $-10.18$ $(\times 1)$ &  $1.91$ $(\times 1)$ &  $3.14$ $(\times 1)$ &  $4.00$ $(\times 1)$  \\
C$_4$H$_{10}$O  &  ethoxy ethane  &  $-12.36$ $(\times 1)$ &  $-11.50$ $(\times 1)$ &  $-9.75$ $(\times 1)$ &  $2.96$ $(\times 1)$ &  $3.46$ $(\times 1)$ &  $3.93$ $(\times 1)$  \\
CH$_2$O$_2$  &  formic acid  &  $-14.94$ $(\times 1)$ &  $-12.55$ $(\times 1)$ &  $-11.42$ $(\times 1)$ &  $2.70$ $(\times 1)$ &  $3.07$ $(\times 1)$ &  $4.29$ $(\times 1)$  \\
HOOH  &  hydrogen peroxide  &  $-15.38$ $(\times 1)$ &  $-12.84$ $(\times 1)$ &  $-11.39$ $(\times 1)$ &  $3.01$ $(\times 1)$ &  $3.02$ $(\times 1)$ &  $4.87$ $(\times 1)$  \\
H$_2$O  &  water  &  $-18.90$ $(\times 1)$ &  $-14.70$ $(\times 1)$ &  $-12.48$ $(\times 1)$ &  $2.88$ $(\times 1)$ &  $4.91$ $(\times 1)$ &  $13.32$ $(\times 1)$  \\
CO$_2$  &  carbon dioxide  &  $-18.11$ $(\times 1)$ &  $-17.99$ $(\times 2)$ &  $-13.73$ $(\times 1)$ &  $2.80$ $(\times 1)$ &  $4.29$ $(\times 2)$ &  $6.59$ $(\times 1)$  \\
CS$_2$  &  carbon disulfide  &  $-14.57$ $(\times 1)$ &  $-13.27$ $(\times 2)$ &  $-10.01$ $(\times 2)$ &  $0.29$ $(\times 2)$ &  $3.34$ $(\times 1)$ &  $4.46$ $(\times 1)$  \\
OCS  &  carbon oxide sulfide  &  $-16.12$ $(\times 1)$ &  $-16.12$ $(\times 2)$ &  $-11.24$ $(\times 2)$ &  $1.85$ $(\times 2)$ &  $3.13$ $(\times 1)$ &  $4.78$ $(\times 1)$  \\
OCSe  &  carbon oxide selenide  &  $-15.89$ $(\times 2)$ &  $-15.63$ $(\times 1)$ &  $-10.50$ $(\times 2)$ &  $1.44$ $(\times 2)$ &  $2.60$ $(\times 1)$ &  $4.18$ $(\times 1)$  \\
CO  &  carbon monoxide  &  $-19.42$ $(\times 1)$ &  $-15.49$ $(\times 1)$ &  $-14.37$ $(\times 1)$ &  $1.22$ $(\times 2)$ &  $5.30$ $(\times 1)$ &  $6.59$ $(\times 1)$  \\
O$_3$  &  ozone  &  $-13.48$ $(\times 1)$ &  $-12.93$ $(\times 1)$ &  $-12.79$ $(\times 1)$ &  $-1.52$ $(\times 1)$ &  $5.23$ $(\times 1)$ &  $7.28$ $(\times 1)$  \\
SO$_2$  &  sulfur dioxide  &  $-13.50$ $(\times 1)$ &  $-13.12$ $(\times 1)$ &  $-12.37$ $(\times 1)$ &  $-0.34$ $(\times 1)$ &  $3.98$ $(\times 1)$ &  $4.39$ $(\times 1)$  \\
BeO  &  beryllium monoxide  &  $-26.77$ $(\times 1)$&  $-10.97$ $(\times 1)$ &  $-9.88$ $(\times 2)$ &  $-2.01$ $(\times 1)$ &  $2.31$ $(\times 1)$ &  $2.48$ $(\times 1)$  \\
MgO  &  magnesium monoxide  &  $-24.89$ $(\times 1)$ &  $-8.76$ $(\times 1)$ &  $-8.17$ $(\times 2)$ &  $-1.29$ $(\times 1)$ &  $1.16$ $(\times 2)$ &  $2.90$ $(\times 1)$  \\
C$_7$H$_8$  &  toluene  &  $-11.74$ $(\times 1)$ &  $-9.19$ $(\times 1)$ &  $-8.90$ $(\times 1)$ &  $1.71$ $(\times 1)$ &  $1.85$ $(\times 1)$ &  $2.97$ $(\times 1)$  \\
C$_8$H$_{10}$  &  ethylbenzene  &  $-11.57$ $(\times 1)$ &  $-9.15$ $(\times 1)$ &  $-8.85$ $(\times 1)$ &  $1.76$ $(\times 1)$ &  $1.76$ $(\times 1)$ &  $2.85$ $(\times 1)$  \\
C$_6$F$_6$  &  hexafluorobenzene  &  $-14.09$ $(\times 1)$ &  $-13.12$ $(\times 1)$ &  $-10.15$ $(\times 2)$ &  $1.08$ $(\times 2)$ &  $1.15$ $(\times 1)$ &  $3.66$ $(\times 2)$  \\
C$_6$H$_5$OH  &  phenol  &  $-11.99$ $(\times 1)$ &  $-9.42$ $(\times 1)$ &  $-8.69$ $(\times 1)$ &  $1.62$ $(\times 1)$ &  $2.35$ $(\times 1)$ &  $2.84$ $(\times 1)$  \\
C$_6$H$_5$NH$_2$  &  aniline  &  $-11.01$ $(\times 1)$ &  $-9.21$ $(\times 1)$ &  $-7.98$ $(\times 1)$ &  $1.83$ $(\times 1)$ &  $2.29$ $(\times 1)$ &  $2.82$ $(\times 1)$  \\
C$_5$H$_5$N  &  pyridine  &  $-10.45$ $(\times 1)$ &  $-9.74$ $(\times 1)$ &  $-9.72$ $(\times 1)$ &  $1.24$ $(\times 1)$ &  $1.62$ $(\times 1)$ &  $3.21$ $(\times 1)$  \\
C$_5$H$_5$N$_5$O  &  guanine  &  $-10.05$ $(\times 1)$ &  $-9.81$ $(\times 1)$ &  $-8.04$ $(\times 1)$ &  $1.57$ $(\times 1)$ &  $1.87$ $(\times 1)$ &  $1.98$ $(\times 1)$  \\
C$_5$H$_5$N$_5$  &  adenine  &  $-9.59$ $(\times 1)$ &  $-9.39$ $(\times 1)$ &  $-8.33$ $(\times 1)$ &  $1.28$ $(\times 1)$ &  $2.06$ $(\times 1)$ &  $2.51$ $(\times 1)$  \\
C$_4$H$_5$N$_3$O  &  cytosine  &  $-9.66$ $(\times 1)$ &  $-9.54$ $(\times 1)$ &  $-8.78$ $(\times 1)$ &  $0.92$ $(\times 1)$ &  $2.29$ $(\times 1)$ &  $2.51$ $(\times 1)$  \\
C$_5$H$_6$N$_2$O$_2$  &  thymine  &  $-10.55$ $(\times 1)$ &  $-10.19$ $(\times 1)$ &  $-9.15$ $(\times 1)$ &  $0.77$ $(\times 1)$ &  $2.15$ $(\times 1)$ &  $2.42$ $(\times 1)$  \\
C$_4$H$_4$N$_2$O$_2$  &  uracil  &  $-10.65$ $(\times 1)$ &  $-10.29$ $(\times 1)$ &  $-9.57$ $(\times 1)$ &  $0.70$ $(\times 1)$ &  $2.10$ $(\times 1)$ &  $2.35$ $(\times 1)$  \\
CH$_4$N$_2$O  &  urea  &  $-10.65$ $(\times 1)$ &  $-10.52$ $(\times 1)$ &  $-10.08$ $(\times 1)$ &  $2.33$ $(\times 1)$ &  $3.51$ $(\times 1)$ &  $4.09$ $(\times 1)$  \\
Ag$_2$  &  silver dimer  &  $-11.08$ $(\times 2)$ &  $-10.82$ $(\times 1)$ &  $-7.41$ $(\times 1)$ &  $-0.70$ $(\times 1)$ &  $1.00$ $(\times 2)$ &  $1.41$ $(\times 1)$  \\
Cu$_2$  &  copper dimer  &  $-9.37$ $(\times 2)$ &  $-9.22$ $(\times 1)$ &  $-7.38$ $(\times 1)$ &  $-0.34$ $(\times 1)$ &  $2.11$ $(\times 1)$ &  $2.16$ $(\times 2)$  \\
NCCu  &  copper cyanide  &  $-12.17$ $(\times 1)$ &  $-11.21$ $(\times 2)$ &  $-10.69$ $(\times 1)$ &  $-0.98$ $(\times 1)$ &  $1.92$ $(\times 2)$ &  $3.08$ $(\times 1)$  \\
\hline\hline
\caption{Quasiparticle energies (negative of the ionization potentials and electron affinities)                 of molecules in the $GW$100 calculated with IP/EA-EOM-CCSD in the def2-TZVPP basis set.}
\label{tab:eomccsd}
\end{longtable*}

\subsection{Accuracy of approximate EOM-CCSD}

We next assess the accuracy of approximations to EOM-CCSD, using the $GW$100
test set.  Henceforth, we compare all approximations to EOM-CCSD, and not to
$\Delta$CCSD(T), for a number of reasons.  First, the comparison is perhaps the
most fair because all approximate techniques are derived from EOM-CCSD, and so
the most we can expect is that they reproduce this parent method.  Second,
although EOM-CCSD was shown above to provide an accurate reproduction of the
$\Delta$CCSD(T) values, the latter approach can be challenging for
open-shell systems like those used in the $(N\pm 1)$-electron
calculations.  For example, while the unrestricted formalism used
in Ref.~\onlinecite{Krause2015} provides a better approximate treatment
of multireference effects, it also suffers from spin contamination,
which can affect the IPs and EAs by up to 0.5~eV, as discussed in
Ref.~\onlinecite{Richard2016}.
Third, it allows us to compare EAs, which are not available in the literature
based on $\Delta$CCSD(T).

\begin{table}[t]
\begin{tabular}{l r r r r}
\hline\hline
Method & IP ME (eV) & IP MAE (eV) & EA ME (eV) & EA MAE \\
\hline
EOM-linCCSD     & $ 0.13$   & $0.14$  & $0.05$  & $0.11$  \\
EOM-CC2         & $ 0.00$   & $0.14$  & $0.08$  & $0.15$ \\
EOM-MBPT2       & $ 0.03$   & $0.13$  & $0.08$  & $0.16$  \\
P-EOM-CC2       & $-0.08$   & $0.12$  & $-0.04$ & $0.08$\\
P-EOM-MBPT2     & $-0.08$   & $0.16$  & $-0.03$ & $0.08$ \\
GF2@HF          & $-0.38$   & $0.42$  & $-0.19$ & $0.22$ \\
$G_0W_0$@HF~\cite{Caruso2016} & $0.26$  & $0.35$ & & \\
$G_0W_0$@PBE~\cite{Caruso2016} & $-0.69$  & $0.69$ & &  \\
\hline\hline
\label{tab:approx}
\end{tabular}
\caption{Mean error (ME) and mean absolute error (MAE) in eV of ionization potentals
(IPs) and electron affinities (EAs)
for molecules contained in the $GW$100 test set.  Error metrics are calculated
with respect to IP/EA-EOM-CCSD without approximation, except for the $GW$ results from
Ref.~\onlinecite{Caruso2016}, which are calculated with respect to $\Delta$CCSD(T)
results from Ref.~\onlinecite{Krause2015}.}
\end{table}

In Tab.~\ref{tab:approx}, we present IP and EA error metrics for a variety of approximate
techniques (all 100 molecules were studied by each approach except for CC2-based
approaches, because the ground-state CC2 failed to converge for twelve molecules).  
Perhaps most remarkably, we find that all CC-based methods exhibit MEs of
less than 0.13~eV and MAEs of less than 0.16~eV.  
Overall, we see that approximations in the ground-state calculation lead to
an average increase in the IP or EA and approximations in the EOM calculation lead
to an average decrease in the IP or EA.  
This behavior can be understood because most perturbative approximations to 
CCSD lead to overcorrelation, which decreases the ground-state energy (increases the IP or EA)
or decreases the excited-state energy (decreases the IP or EA).

Without any partitioning, the EOM-linCCSD, EOM-CC2, and EOM-MBPT2 all perform
similarly.  Although the error incurred by the most expensive linearized CCSD is
slightly larger (compared to the other approximate treamtents of the ground
state), the error is extremely systematic with a very small spread; for example,
over 50 molecules overestimate the IP by 0.1~eV and another 30 molecules
overestimate the IP by 0.2~eV.  
All $N^5$ approximate methods -- based on CC2 or MBPT2, with or without
partitioning -- perform impressively well.  EOM-CC2 and EOM-MBPT2 exhibit very
similar results (even on the level of individual molecules), suggesting that the
orbital relaxation due to $\hat{T}_1$ is not important in many of these cases.
With partitioning, the ME becomes only slightly negative without any significant
increase in the MAE.

The qualitatively similar performance of all approximate EOM-CC methods and their
improvement compared to $G_0W_0$@HF suggests
that the precise details of screening are not important in molecules, and perhaps
second-order exchange is more important.  To test this, we also show the results
of IPs and EAs calculated using the second-order Green's function (GF2); the mean error is large
and negative, $-0.38$~eV for IPs and $-0.19$~eV for EAs, which can be compared
to the IP results of $G_0W_0$@HF, $+0.26$~eV.  Remarkably, the
P-EOM-MBPT2 approach, which has essentially identical cost as GF2, reduces the
mean error of the latter to only $-0.08$~eV.  This result suggests that the
combination of second-order exchange with a small amount of screening, beyond
the second-order ring diagram, is important for quantitative accuracy.  Given
the extremely low cost, we identify P-EOM-MBPT2 as an attractive low-cost
approach for IPs and EAs of larger molecules, with potential applications in the
solid state. However, it must be kept in mind that the increased importance of
screening in solids may preclude the success of perturbative approximations. 

\section{Conclusions and outlook}
\label{sec:conc}

To summarize, we have presented a diagrammatic, algebraic, and numerical evaluation
of quasiparticle excitation energies predicted by EOM-CCSD, especially
as compared to those of the $GW$ approximation.  
Although the EOM-CCSD Green's function includes fewer ring diagrams than the
$GW$ approximation, we find that its inclusion of many more diagrams --
including ladders and exchange -- produces excitation energies that are much
more accurate than those from the $GW$ approximation.
To completely encompass all $GW$ diagrams requires the use of non-perturbative
EOM-CCSDT.  

We also investigated the accuracy of a number of cost-saving approximations to
EOM-CCSD, many of which reduce the canonical scaling to $N^5$ (which could be
further reduced through density-fitting~\cite{Schutz2003} or tensor
hypercontraction~\cite{Hohenstein2013}).  All CC-based approximations considered
yield very small errors on average.  For systems where screening is relatively
unimportant, such as molecules or large band-gap insulators, we identify
P-EOM-MBPT2 as an accurate and inexpensive $N^5$ approach.  We attribute the
success of P-EOM-MBPT2 to its exact treatment of screening and exchange through
second order (as in GF2), combined with a small number of third-order diagrams. 

We anticipate that the framework and connections laid out here will aid 
future work on the $GW$ approximation, through the identification of the most
important excluded diagrams.
With respect to IP/EA-EOM-CCSD calculations of band structures in solids~\cite{McClain2017},
the present work motivates efforts to quantify the error induced by neglecting
some of the non-TDA ring diagrams, which are conventionally thought to be crucial
for screening in solids. In the same vein, the inclusion of triple excitations,
perhaps even perturbatively, could be an important ingredient in recovering --
and rigorously surpassing -- RPA physics.

\section*{Acknowledgments}
T.C.B.~thanks Garnet Chan for helpful conversations related to this work,
Marco Govoni for discussions about the $GW$100, and Wim Klopper for sharing the
data from Ref.~\onlinecite{Krause2015}.
We thank Alan Lewis for a critical reading of the manuscript.
Diagrams were drawn using the JaxoDraw program~\cite{Binosi2004}.
This work was supported by start-up funding from the University of Chicago and
by the Midwest Integrated Center for Computational Materials (MICCoM), as part
of the Computational Materials Sciences Program funded by the U.S.  Department
of Energy (DOE), Office of Science, Basic Energy Sciences, Materials Sciences
and Engineering Division (5J-30161-0010A).

\end{document}